\documentclass[8pt]{revtex4}
\pdfoutput=1
\usepackage{amssymb}
\usepackage{latexsym}
\usepackage{epsfig}

\begin{document}
\title{{\bf  A mathematical model for  DNA }}

\author{  Alireza Sepehri $^{1}$ \footnote{
		email address: alireza.sepehri2017@gmail.com} }
\address{$^1$ Research Institute for Astronomy and Astrophysics of Maragha
	(RIAAM), Maragha, Iran,P.O.Box:55134-441.\\ $^2$ Research
	Institute for biotech development, Tehran, Iran.}

\begin{abstract}
Recently, some authors have shown that a DNA molecule produces electromagnetic signals and  communicates with other DNA molecules or other molecules. In fact, a DNA acts like a receiver or transmitter of radio waves. In this paper, we  suggest a mathematical model for the DNA molecule and use of its communication to cure some diseases like cancer. In this model, first, by using concepts from string theory and M-theory, we  calculate the energy of a  DNA in terms of interactions between free electrons and bound electrons.  We show that when a DNA is damaged,  its energy changes and an extra  current  is produced. This extra current causes the electromagnetic signals of a damaged DNA molecule to be different when compared to the electromagnetic signals of a normal DNA molecule. The electromagnetic signals of a damaged DNA molecule induces an extra current in a normal DNA molecule and leads to its destruction. By sending crafted electromagnetic signals to normal DNA molecules and inducing an opposite current with respect to this extra current, we can prevent the destruction of normal DNA. Finally, we argue  that the type of packing of DNA in chromosomes of men and women are different. This causes radiated waves from DNAs of men and women to have opposite signs and  cancel the effect of each other in a pair.  Using this property, we suggest another mechanism to cancel the effect of extra waves, which are produced by  DNAs in cancer cells of a male or a female, by extra waves which are produced by  DNAs in similar cells of a female or a male and prevent the progression of the disease. We also discuss about the radiated waves from sperms, eggs, teleportation of sperms into eggs and introduce new Virus Medical Imaging Technique .  Finally, we will argue about the existence of  information coding and gender of water  and  waves and introduce new wave topoisomerase in 4 + n-dimensions where n is the number of extra dimensions. We also discuss about two DNA-like structures of water and waves in extra dimensions. We will show that in some conditions, topoisomerase-like waves could extract the structure of DNA from pure water in extra dimension and create life.  Using the stored  topoisomerase-like waves in some objects in nature like trees, we could control the growth of cells of  fetus of chicken, animal and others without needing to the egg and produce new born before normal time. We can produce a fetus by returning specialized cells of a hen (Chicken) and a rooster to stem ones and then mixing them in material of an egg in some thermal conditions. 

 \vspace{5mm}\noindent\vspace{0.8mm}\newline Keywords: DNA, Mathematical model, Teleportation, Cancer, Electromagnetic Signal
\end{abstract}

\maketitle
\section{Introduction}

The electronic properties of a DNA  molecule is one of main subjects in biophysics, which occupies many researchers. This is because that each type of defects or damage to a DNA has a direct effect on its electronic properties. In the past, there have been many studies on this subject. For example, some authors have argued about the charge transport in a single-stranded DNA molecule in the direction perpendicular to the backbone axis and discussed that the different electronic  structure of the four bases causes a difference in conductance of the nucleotides \cite{e1}. In another paper, the electronic transfer properties of a DNA molecule have been investigated in terms of localization lengths and  a novel enhancement of localization lengths have been obtained at some energies for an progressing binary backbone disorder \cite{e2}. In another research effort, authors have discussed the possibility of a mechanism for fast sequencing based on the distributions of transverse electrical currents in single-stranded DNA. Their method gives this opportunity  to study, predict and cure diseases from the perspective of the genetic compound of each individual \cite{e3}. In yet another study, authors have observed the high spin selectivity and the length-dependent spin polarization in double-stranded DNA \cite{e4}. Also, they have shown that  the spin polarization in double-stranded DNA is significant even in the case of weak spin-orbit coupling, while no spin polarization can be seen in single-stranded DNA \cite{e5}. Now, the question arises: could we consider the evolution of DNA by detecting and analyzing their electromagnetic radiation?

In 2009, Montagnier and his collaborators have investigated about the capacity of some bacterial DNA sequences to radiate very low frequency electromagnetic signals in high aqueous dilutions. They have argued that the genomic DNA of most pathogenic bacteria contains sequences which are able to produce such signals \cite{DNA1}. In another research, authors have explained the experimental conditions by which electromagnetic waves (EMS) of low frequency can be radiated by diluted aqueous solutions of some bacterial and viral DNAs. Also, they have detected this  transduction process in living human cells exposed to EMS irradiation and proposed a quantum field theory analysis of the phenomenon \cite{DNA2}. In parallel investigations, some other authors have considered the responses of DNA to electromagnetic fields in different frequency ranges, and specified the properties of DNA molecules as antennas \cite{DNA3}. Motivated by such research, in this paper we suggest a mathematical model for DNA and obtain its the energy and the current  using the concepts of M-theory.  We design an electrical circuit for communication with DNA and for the prevention of disease progression. We will show that by analyzing the radiated waves from DNA molecules, we can consider their development and defects. Then, by sending electromagnetic signals to DNA molecules, we can cancel the effect of  damaged DNAs on the normal DNAs.

The outline of this paper is as  follows. In section \ref{oo1},  we propose a mathematical model for calculating the energy of bases in a DNA molecule. In section \ref{o1}, we design a model for the electrical circuit of a normal DNA. In section \ref{o2}, we study the origin of the radiated waves from a damaged DNA and suggest a model for preventing disease progression. In section \ref{o3}, we suggest a mechanism to use the difference between radiated waves from DNAs in men and women to cure the cancer.  The last section is devoted to summary and conclusions.

\section{A mathematical model for calculating the energy of bases in a DNA }\label{oo1}
In this section, first, we introduce polygonal molecules and construct an $N$-dimensional polygonal manifold. To this aim, we use concepts of string theory. We assume that electrons are similar to strings and the hexagonal shape of molecules is similar to the shape of polygonal manifolds. We will show that at first stage, there are molecules (point-like polygonal manifolds) in the human body (see Figure 1) and electrons (strings) are attached to them. Polygonal manifolds are manifolds with a $q$-gonal shape which are similar to molecules of DNA. These manifolds have only one dimension, in the direction of time. All interactions between electrons are simulated by interactions between  strings. The potential of these interactions can be given by a delta function and thus, the energy of manifold shrinks to one.  We can write:

\begin{eqnarray}
&& V(\sqrt{C(\tilde{X}^{I})})= \delta(\sqrt{C(\tilde{X}^{I})}) \nonumber\\&&  E_{M^{0}}=1=\int_{M^{0}} d \sqrt{C(\tilde{X}^{I})} V(\sqrt{C(\tilde{X}^{I})})=\int_{M^{0}} d\sqrt{C(\tilde{X}^{I})} \delta(\sqrt{C(\tilde{X}^{I})})= \nonumber\\&&  \int_{M^{0}} d \sqrt{C(\tilde{X}^{I})} e^{-\sqrt{C(\tilde{X}^{I})}\sqrt{C(\tilde{X}_{I})}}	\label{spm1}
\end{eqnarray}

where $M^{0}$  represents the point-like manifold, $\tilde{X}^{I}$'s are strings hat are attached to them and $y$ is the length of the point, which shrinks to zero. Also, $C(\tilde{X}^{I})$ is the equation of the polygonal manifold. In the above equation, $C(\tilde{X}^{I})$ has the following relation with $\tilde{X}^{I}_{i}$:

\begin{eqnarray}
&& C(\tilde{X}^{I})=\Sigma_{i=1}^{q} \frac{(\tilde{X}^{I}_{i})^{2}}{2\pi y \tilde{a}_{i}^{2}}	\label{ww1}
\end{eqnarray}

where $\tilde{a}_{i}$ is the coefficient of the $i$-th side of a point-like polygonal manifold. For example, for a sphere, we can write:

\begin{eqnarray}
&& C(\tilde{X}^{I})= \frac{(\tilde{X}^{I}_{1})^{2}}{2\pi y R^{3}} +  \frac{(\tilde{X}^{I}_{2})^{2}}{2\pi y R^{3}} + \frac{(\tilde{X}^{I}_{3})^{2}}{2\pi y R^{3}}	\label{ww2}
\end{eqnarray}

Using a new redefinition of string fields  $\tilde{X}^{I}\longrightarrow \sqrt{2\pi y} X^{I}$, we obtain:

\begin{eqnarray}
&& E_{M^{0}}=1= \int_{M^{0}} d \sqrt{C(X^{I})} e^{-\pi \sqrt{C(X^{I})}\sqrt{C(X_{I})}}	\label{spm2}
\end{eqnarray}

By calculating the integral, we can  obtain a solution for strings ($X^{I}$):

\begin{eqnarray}
&& \int_{M^{0}} d\sqrt{C(X^{I})} e^{-\pi \sqrt{C(X^{I})}\sqrt{C(X_{I})}}=1\longrightarrow \nonumber\\&&  \frac{1}{2}erf(\sqrt{ \sqrt{C(X^{I})}\sqrt{C(X_{I})}\pi})=1 \longrightarrow \nonumber\\&& C(X^{I})\approx I\label{spm3}
\end{eqnarray}

where $I$ is a unitary matrix. This equation shows that at the first stage, there is no interaction between strings and they are the same. In fact, there is a very high symmetry for the early stages of system and all matters have the same origin. For example, for a sphere, the above equation yields:

\begin{eqnarray}
&& \frac{(X^{I}_{1})^{2}}{R^{3}} +  \frac{(X^{I}_{2})^{2}}{R^{3}} + \frac{(X^{I}_{3})^{2}}{R^{3}}\approx I\label{ww3}
\end{eqnarray}

The above example shows that initial strings constitute a spherical manifold with radius $R$. By changing $\tilde{a}_{i}$ and number of sides, the shape of a point-like manifold can be changed.

\begin{figure*}[thbp]
	\begin{center}
		\begin{tabular}{rl}
			\includegraphics[width=5cm]{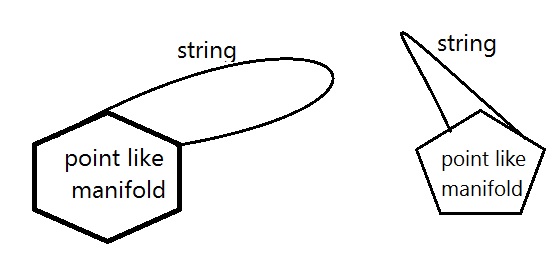}
		\end{tabular}
	\end{center}
	\caption{ One example of point-like manifolds and attached strings. }
\end{figure*}

We can extend this mechanism to $N$-dimensional manifolds (see Figure 2). Each $N$-dimensional manifold is similar to an $N$-dimensional DNA.  This manifold can be built by joining  $(N+1)$ point-like manifolds. The potential of interactions between strings on them can be described by a delta function. In fact, all interactions on a point-like manifold are concentrated at a point and thus the total potential is approximately infinite. This potential, which is zero at other points and infinite at an special point, can be represented by a delta function. Integrating over all these potentials, we can obtain the total energy of system:

\begin{eqnarray}
&& E_{M_{i}^{0}}=\int_{M_{i}^{0}} d C (\tilde{ X_{i}}^{I}) V( C(\tilde{X_{i}}^{I}))=\int_{M_{i}^{0}} d C(\tilde{X_{i}}^{I}) \delta(C (\tilde{X_{i}}^{I}))=1\Rightarrow \nonumber\\&&\nonumber\\&& \nonumber\\&& E_{M_{1}^{0}+...+M_{N+1}^{0}}=1=\int_{M_{1}^{0}+...+M_{N+1}^{0}} d C(\tilde{X_{1}}^{I})...d C(\tilde{X_{N+1}}^{I}) \delta(C(\tilde{X_{1}}^{I}))...\delta(C (\tilde{X_{N+1}}^{I}))=\nonumber\\&&  \int_{M_{1}^{0}+...+M_{N+1}^{0}} d C(\tilde{X_{1}}^{I})...d C(\tilde{X_{N+1}}^{I})e^{-C(\tilde{X_{1}}^{I}) C(\tilde{X_{1}}_{I})}...e^{-C(\tilde{X_{N+1}}^{I}) C(\tilde{X_{N+1}}_{I})}\nonumber\\&&	\label{spm24}
\end{eqnarray}

\begin{figure*}[thbp]
	\begin{center}
		\begin{tabular}{rl}
			\includegraphics[width=5cm]{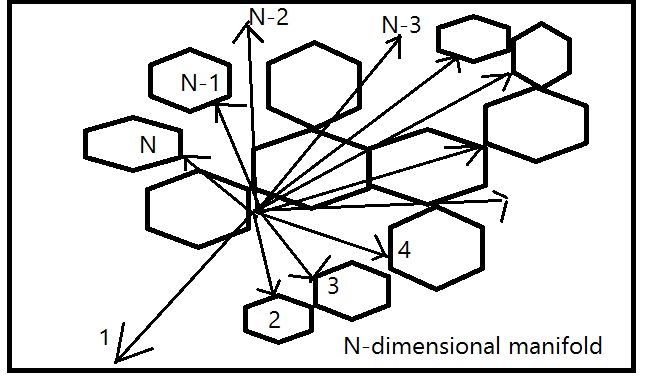}
		\end{tabular}
	\end{center}
	\caption{  Constructing an $N$-dimensional manifold by joining point-like manifolds. }
\end{figure*}

In the above equation, the energy of the manifold is normalized to one. It is concluded that for an $N$-dimensional manifold, there are $N+1$ integrations. An extra dimension corresponds to time and in totality, we have an $(N+1)$-dimensional manifold. Using a new definition of  the string fields  $\tilde{X}^{I}\longrightarrow \sqrt{2\pi y} X^{I}$, we can obtain:

\begin{eqnarray}
&& E_{M^{N}}=1= \int_{M_{1}^{0}+....+M_{N+1}^{0}} d \sqrt{C (X_{1}^{I})}... d \sqrt{C(X_{N+1}^{I})} e^{-\pi \sqrt{C (X_{1}^{I})}\sqrt{C (X_{1 I})}}....e^{-\pi \sqrt{C( X_{N+1}^{I})}\sqrt{C (X_{N+1 I})}}	\label{spm25}
\end{eqnarray}

By joining  point-like manifolds and building an $N$-dimensional manifold,  strings will be  functions of coordinates of $(N+1)$-manifolds ($X^{I}(y_{1}....y_{N+1})$). Thus, equation (\ref{spm25}) can be written in the form

\begin{eqnarray}
&& E_{M^{N}}=1=  \int_{M_{1}^{0}+...+M_{N+1}^{0}}dy_{1}^{I}...dy_{N+1}^{I} (\Sigma_{i_{1},...i_{N+1}=1...N+1}\frac{d \sqrt{C (X_{i_{2}}^{I})}}{dy_{i_{1}}^{I}}\frac{d \sqrt{C(X_{i_{1}}^{I})}}{dy_{i_{2}}^{I}}...\frac{d \sqrt{C (X_{i_{N+1}}^{I})}}{dy_{i_{N}}^{I}}\frac{d \sqrt{C(X_{i_{N}}^{I})}}{dy_{i_{N+1}}^{I}} \nonumber\\&& e^{-\pi \sqrt{C(X_{i_{1}}^{I})} \sqrt{C (X_{i_{1} I})}}...e^{-\pi \sqrt{C(X_{i_{N+1}}^{I})} \sqrt{C(X_{i_{N+1} I})}} +	\nonumber\\&& \int_{M_{1}^{0}+...+M_{N+1}^{0}}dy_{i_{1}}^{I}...dy^{i_{1}I} (\frac{d \sqrt{C (X_{i_{1}}^{I})}}{dy_{i_{1}}^{I}}....\frac{d \sqrt{C(X_{i_{N+1} I})}}{dy_{i_{1} I}}) e^{-\pi \sqrt{C (X_{i_{1}}^{I}}) \sqrt{C (X_{i_{1} I})}}...e^{-\pi \sqrt{C (X_{i_{N+1}}^{I})} \sqrt{C (X_{i_{N+1} I})}}+\nonumber\\&&...+\nonumber\\&&  \int_{M_{1}^{0}+...+M_{N+1}^{0}}dy_{i_{N+1}}^{I}...dy^{i_{N+1}I} (\frac{d \sqrt{C (X_{i_{1}}^{I})}}{dy_{i_{N+1}}^{I}}....\frac{d \sqrt{C (X_{i_{N+1} I})}}{dy_{i_{N+1} I}}) e^{-\pi\sqrt{ C (X_{i_{1}}^{I})} \sqrt{C (X_{i_{1} I})}}...e^{-\pi \sqrt{C (X_{i_{N+1}}^{I})} \sqrt{C (X_{i_{N+1} I})}}\label{spm26}
\end{eqnarray}

Taylor-expanding the exponential functions over the crossed points, we obtain

\begin{eqnarray}
&& E_{M^{N}}=1=
\int_{M_{1}^{0}+...+M_{N+1}^{0}}dy_{1}^{I}...dy_{N+1}^{I} (\Sigma_{i_{1},...i_{N+1}=1...N+1}\frac{d \sqrt{C (X_{i_{2}}^{I})}}{dy_{i_{1}}^{I}}\frac{d \sqrt{C (X_{i_{1}}^{I})}}{dy_{i_{2}}^{I}}...\frac{d \sqrt{C(X_{i_{N+1}}^{I})}}{dy_{i_{N}}^{I}}\frac{d \sqrt{C(X_{i_{N}}^{I})}}{dy_{i_{N+1}}^{I}}\times \nonumber\\&& \pi^{N+1} (\Sigma_{i_{1},...,i_{N+1}=1,,,,N+1}(\frac{\partial}{\partial y_{i_{1} I}}(\sqrt{C (X_{i_{1}}^{I})}\sqrt{ C (X_{i_{1} I})})+....+\frac{\partial}{\partial y_{i_{1} I}}...\frac{\partial}{\partial y_{i_{N+1}}^{I}}(\sqrt{C (X_{i_{1}}^{I}}) \sqrt{C (X_{i_{1} I})})+1)\times \nonumber\\&&....\times \nonumber\\&&(\frac{\partial}{\partial y_{i_{1} I}}(\sqrt{C (X_{i_{N+1}}^{I})} \sqrt{C (X_{i_{N+1} I})})+...+\frac{\partial}{\partial y_{i_{1} I}}...\frac{\partial}{\partial y_{i_{N+1}}^{I}}(\sqrt{C (X_{i_{N+1}}^{I})} \sqrt{C (X_{i_{N+1} I})})+1)\times \nonumber\\&&\Big(\Sigma_{i_{1},...i_{N+1}=0...N+1}(y_{i_{N+1}}^{I}-y_{0}^{I})^{i_{N+1}}....(y_{1 I}-y_{0 I})^{i_{1}}\Big) +	 ...\label{spm27}
\end{eqnarray}

We now wish to obtain the relation between different derivatives of strings respect to coordinates and various types of the matter, such as scalars, fermions, gauge fields and gravitons. For this, we compare the action of branes in string theory with the action of matter in field theory.  We can write \cite{D3}:

\begin{eqnarray}
S_{Gravity-Matter}&=& S_{D3}   \label{D4}
\end{eqnarray}

where

\begin{eqnarray}
S_{D3}&=& -T_{D3}\int d^{4}y \sqrt{1+ g_{ij}\partial_{a}X^{i}\partial^{a}X^{j} -4\pi^{2} l_{s}^{4}F_{ab}F^{ab}},\nonumber\\ S_{Gravity-Matter}&=& \int d^{4}y \sqrt{-g}\Big(R +g_{ab}\partial^{a}\phi\partial^{b}\phi -i\bar{\psi}\gamma^{a}\partial_{a}\psi+A_{i}A^{i}+\frac{1}{2}\phi^{2}+1\Big),  \label{D2}
\end{eqnarray}

Here, $A_{b}$ is the gauge field, $F_{ab}$ is the field strength, $X^{\mu}$ is the string, $g_{\mu\nu}$ is the metric, $T_{D3}$ is tension and $l_{s}$ is the string length. Also, $\phi$ is the scalar field and $\psi$ is the fermionic field. Using equation (\ref{D4}), we can write the following relations between strings and fields (see Figures 3,4 and 5):

\begin{figure*}[thbp]
	\begin{center}
		\begin{tabular}{rl}
			\includegraphics[width=5cm]{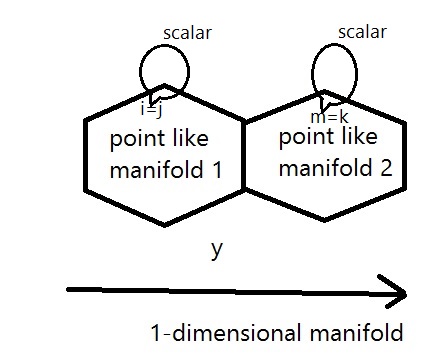}
		\end{tabular}
	\end{center}
	\caption{  Strings with both ends located on a point-like manifold form scalar fields ($\phi$). }
\end{figure*}

\begin{figure*}[thbp]
	\begin{center}
		\begin{tabular}{rl}
			\includegraphics[width=5cm]{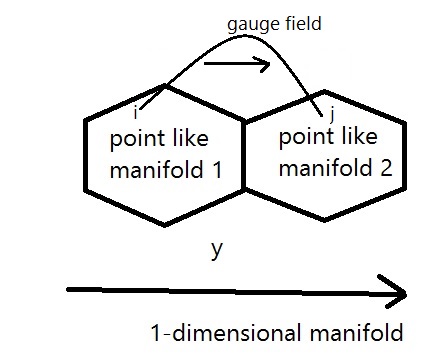}
		\end{tabular}
	\end{center}
	\caption{  Strings with one end on a point-like manifold and the other on another point-like manifold form gauge fields ($F$-fields ($F_{ij}$)). }
\end{figure*}

\begin{figure*}[thbp]
	\begin{center}
		\begin{tabular}{rl}
			\includegraphics[width=5cm]{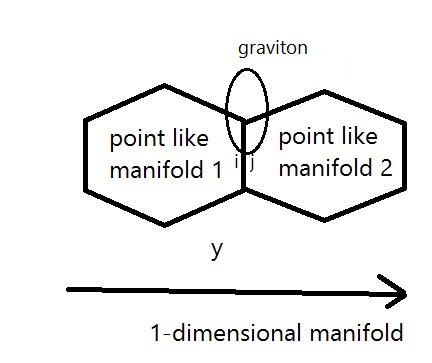}
		\end{tabular}
	\end{center}
	\caption{  Strings with two ends near a linked point, yielding a symmetric shape, form graviton fields ($g_{ij}$). }
\end{figure*}

\begin{eqnarray}
&& X_{1}^{I}X_{1 I}=....= X_{j}^{I}X_{j I} \nonumber\\&& \nonumber\\&&  \frac{dX_{i}^{I}}{dy_{j}^{I}}=e^{i}_{j} \quad i,j=1,2 \quad F_{12}\rightarrow  \Big( \frac{\partial}{\partial y_{2 I}}\frac{\partial}{\partial y_{1}^{I}}(X_{1}^{I}X_{1 I})- \frac{\partial}{\partial y_{1 I}}\frac{\partial}{\partial y_{2}^{I}}(X_{1}^{I}X_{1 I})\Big) \nonumber\\&& g_{12}\rightarrow  \Big( \frac{\partial}{\partial y_{2 I}}\frac{\partial}{\partial y_{1}^{I}}(X_{1}^{I}X_{1 I})+ \frac{\partial}{\partial y_{1 I}}\frac{\partial}{\partial y_{2}^{I}}(X_{1}^{I}X_{1 I})\Big)  \nonumber\\&&\nonumber\\&& \phi \rightarrow \frac{\partial}{\partial y_{1 I}}\frac{\partial}{\partial y_{1}^{I}}(X_{1}^{I}X_{1 I})\quad \sqrt{-g}g^{12} \rightarrow (\frac{dX_{1}^{I}}{dy_{2}^{I}}\frac{dX_{2 I}}{dy_{2 I}}) \quad \phi\rightarrow \psi_{\downarrow}\psi_{\uparrow}  \nonumber\\&& \nonumber\\&&       \Sigma_{ijkf}\frac{\partial}{\partial y_{f}^{I}}\frac{\partial}{\partial y_{k I}}\frac{\partial}{\partial y_{j I}}\frac{\partial}{\partial y_{i}^{I}}(X_{i}^{I}X_{i I})\approx  \frac{1}{2}R_{ijkf} +  \epsilon^{ijkf}\frac{\partial}{\partial y_{f}^{I}}\frac{\partial}{\partial y_{k I}}(F_{ij}))\label{spm28}
\end{eqnarray}

We suppose that point-like manifolds are very close to each other and thus, we can obtain:

\begin{eqnarray}
&&  y_{i}^{I}-y_{0}^{I}= ...y_{3}^{I}-y_{0}^{I}= y_{2}^{I}-y_{0}^{I}=y_{1}^{I}-y_{0}^{I}=\sigma\longrightarrow \frac{1}{\pi}  \label{spmtt16}
\end{eqnarray}

Using equations ( \ref{spm28} and  \ref{spmtt16} ) and equation (\ref{spm27}), we obtain:

\begin{eqnarray}
&& E_{M^{N}}=1=  \int_{M^{N}}d^{N+1}y\sqrt{-g} [\Pi_{n'=1}^{N} \Sigma_{\alpha_{n'}=1}^{q_{n'}}\frac{1}{ \tilde{a}_{\alpha_{n'}}^{2}}\Big(R^{\alpha_{n'}}-\frac{1}{2}\partial_{i}\phi^{\alpha_{n'}}\partial^{i}\phi^{\alpha_{n'}} - \frac{1}{4}\epsilon^{ijkm} F_{ij}^{\alpha_{n'}}F_{km}^{\alpha_{n'}}\nonumber\\&&+i\gamma^{i}\bar{\psi}\partial_{i}\psi.....-\nonumber\\&&\frac{1}{N(N-1)..1}\epsilon^{i_{1}i_{2}...i_{N}}( F_{i_{1}i_{2}}^{\alpha_{n'}}...F_{i_{N-3}i_{N-2}}^{\alpha_{n'}}F_{i_{N-1}i_{N}}^{\alpha_{n'}})...  + \frac{1}{N(N-1)..1}\epsilon^{i_{1}i_{2}...i_{N}} (R_{i_{1}i_{2}}^{\alpha_{n'}}...R_{i_{N-3}i_{N-2}}^{\alpha_{n'}}R_{i_{N-1}i_{N}}^{\alpha_{n'}})-\nonumber\\&&.....-\nonumber\\&&\frac{1}{N(N-1)..1}\epsilon^{i_{1}i_{2}...i_{N}}( \partial_{i_{1}}...\partial_{i_{N-2}}F_{i_{N-1}i_{N}}^{\alpha_{n'}})...  + \frac{1}{N(N-1)..1}\epsilon^{i_{1}i_{2}...i_{N}} (\partial_{i_{1}}...\partial_{i_{N-2}}R_{i_{N-1}i_{N}}^{\alpha_{n'}})+ V(\phi^{\alpha_{n'}}) +... \Big)] \nonumber\\&& V(\phi)=1+ \frac{\phi^{2}}{2}..+\frac{\phi^{N}}{N(N-1)...1}  \label{spm29}
\end{eqnarray}

In a DNA molecule, $N$ is the number of bases, $q_{n'}$ is the number of corners in each base, $F$ is the field strengths of photons which are exchanged between bases and $R$ is the curvature of gravity which is exchanged between DNA molecules. This equation shows that by joining point-like polygonal manifolds and building $N$-dimensional ones, the related energy in usual quantum field theory is produced. This energy contains all interactions that may be produced between bases in a DNA molecule.

This energy is seemed to be very distant of results of refs \cite{R7,R8}. However, if we use of the relation between gauge fields, curvatures, scalars and fermions in equation (\ref{spm29}), we can achieve the same results. To this end, we generalize our calculations to M-theory and use of the following replacements for gauge fields and scalars \cite{R7,R8}:

\begin{eqnarray}
&&F_{ab} \text { in string theory} \rightarrow F_{abc} \text { in M-theory},\nonumber\\ && F_{abc}=\partial_{a} A_{bc}-\partial_{b} A_{ca}+\partial_{c}
A_{ab}.\nonumber\\ &&\partial_{a}\phi\partial^{a}\phi \text { in string theory} \rightarrow \partial_{b}\partial_{a}\phi\partial^{a}\partial^{b}\phi \text { in M-theory}\label{wj2}
\end{eqnarray}

These gauge fields and also scalars have a direct relation with spinor fields like electrons \cite{R7,R8}:

\begin{eqnarray}
&& A_{ab}\rightarrow \psi^{U}_{a}\psi^{L}_{b}-\psi^{L}_{a}\psi^{U}_{b}\nonumber \\
&& \phi \rightarrow \psi^{U}_{a}A^{ab}\psi^{L}_{b}-\psi^{L}_{a}A^{ab}\psi^{U}_{b}+\Psi^{U}_{a}A^{ab}\Psi^{L}_{b}-\Psi^{L}_{a}A^{ab}\Psi^{U}_{b}+\Psi^{U}_{a}A^{ab}\psi^{L}_{b}-\psi^{L}_{a}A^{ab}\Psi^{U}_{b}\nonumber \\
&&\partial_{a} =\partial_{a}^{U}+\partial_{a}^{L} \nonumber \\
&&\partial_{a}^{U}\psi^{U}_{a}=1,\;\partial_{a}^{L}\psi^{L}_{a}=1\label{wj4}
\end{eqnarray}

where the indices $L$ and $U$ refer to lower and upper indices. On the other hand, gauge fields and spinors have a directional relation with curvature:

 \begin{eqnarray}
&&A^{ab}=g^{ab}=h^{ab}=h_{1}^{ab}\otimes h_{2}^{ab}-h_{2}^{ab}\otimes h_{1}^{ab} ~ ~  and  ~ ~ a,b,c=\mu,\nu,\lambda
\Rightarrow \nonumber\\&& F_{abc}=\partial_{a} A_{bc}-\partial_{b}
A_{ca}+\partial_{c}
A_{ab}=2(\partial_{\mu}g_{\nu\lambda}+\partial_{\nu}g_{\mu\lambda}-\partial_{\lambda}g_{\mu\nu})=
2\Gamma_{\mu\nu\lambda}\nonumber\\&&\nonumber\\&&\langle
F^{\rho}\smallskip_{\sigma\lambda},F^{\lambda}\smallskip_{\mu\nu}\rangle=\nonumber\\&&\partial_{\nu}
\Gamma^{\rho}_{\sigma\mu}-\partial_{\mu}\Gamma^{\rho}_{\sigma\nu}+\Gamma^{\rho}_{\lambda\nu}
\Gamma^{\lambda}_{\sigma\mu}-\Gamma^{\rho}_{\lambda\mu}\Gamma^{\lambda}_{\sigma\nu}
=R^{\rho}_{\sigma\mu\nu}\nonumber\\&&\langle
F_{abc},F_{a'}^{bc}\rangle=R_{aa'}^{anti-parallel}-R_{aa'}^{parallel}\nonumber\\&& R_{MN}=R_{aa'}+R_{ia'}+R_{ij'}=R_{Free-Free}^{anti-parallel}+R_{Free-Bound}^{anti-parallel}+R_{Bound-Bound}^{anti-parallel}-\nonumber\\&&R_{Free-Free}^{parallel}-R_{Free-Bound}^{parallel}-R_{Bound-Bound}^{parallel}\label{wj7}
\end{eqnarray}

and
\begin{eqnarray}
&&\langle \partial^{b}\partial^{a}\phi,\partial_{b}\partial_{a}\phi\rangle = \Big(\Psi^{\dag a,U}R_{aa'}^{anti-parallel} \Psi^{a',L}+\Psi^{\dag a,L}R_{aa'}^{anti-parallel}\Psi^{a',U}-\nonumber \\
&&\Psi^{\dag a,U}R_{aa'}^{parallel}\Psi^{a',U}-\Psi^{\dag a,L}R_{aa'}^{parallel} \Psi^{a',L}+\nonumber \\
&&\Psi^{\dag a,L}\Psi^{\dag d,U}\partial_{d}\partial^{d'}(R_{aa'}^{parallel}+R_{aa'}^{anti-parallel})\Psi^{a',L}\Psi^{U}_{d'}-\nonumber \\
&&\Psi^{\dag a,L}\Psi^{\dag d,U}\partial_{d}(R_{aa'}^{parallel}+R_{aa'}^{anti-parallel}) \Psi^{a',L}-\nonumber \\
&&\Psi^{\dag a,U}\Psi^{\dag d,L}\partial_{d}(R_{aa'}^{parallel}+R_{aa'}^{anti-parallel})\Psi^{a',U}\Big)+\nonumber \\
&&  \Big(\psi^{\dag i,U}R_{ii'}^{anti-parallel} \psi^{i',L}+\psi^{\dag i,L}R_{ii'}^{anti-parallel}\psi^{i',U}-\nonumber \\
&&\Big(\psi^{\dag i,U}R_{ii'}^{parallel}\psi^{i',U}-\psi^{\dag i,L}R_{ii'}^{parallel}\psi^{i',L}+\nonumber \\
&&\psi^{\dag i,L}\psi^{\dag m,U}\partial_{m}\partial^{m'}(R_{ij'}^{parallel}+R_{ij'}^{anti-parallel})\psi^{i',L}\psi^{U}_{m'}-\nonumber \\
&&\psi^{\dag i,L}\psi^{\dag m,U}\partial_{m}(R_{ij'}^{parallel}+R_{ij'}^{anti-parallel})\psi^{i',L}-\nonumber \\
&&\psi^{\dag i,U}\psi^{\dag m,L}\partial_{m}(R_{ij'}^{parallel}+R_{ij'}^{anti-parallel})\psi^{i',U}\Big)+\nonumber \\
&& \Big(\Psi^{\dag a,U}R_{ai'}^{anti-parallel}\psi^{i',L}+\Psi^{\dag a,L}R_{ai'}^{anti-parallel}\rangle\psi^{i',U}-\nonumber \\
&&\Psi^{\dag a,U}R_{ai'}^{parallel}\psi^{i',U}-\Psi^{\dag a,L}R_{ai'}^{parallel}\psi^{i',L}+\nonumber \\
&&\Psi^{\dag a,L}\Psi^{\dag d,U}\partial_{d}\partial^{i'}(R_{ai'}^{parallel}+R_{ai'}^{anti-parallel})\psi_{j'}^{L}\psi^{i',U}-\nonumber \\
&&\Psi^{\dag a,L}\Psi^{\dag d,U}\partial_{d}(R_{ai'}^{parallel}+R_{ai'}^{anti-parallel})\psi^{i',L}-\nonumber \\
&&\Psi^{\dag a,U}\Psi^{\dag d,L}\partial_{d}(R_{ai'}^{parallel}+R_{ai'}^{anti-parallel})\psi^{i',U}\Big)\approx \nonumber \\
&&( R_{Free-Free}^{parallel})^{2}+( R_{Free-Free}^{anti-parallel})^{2}+( R_{Free-Bound}^{parallel})^{2}+\nonumber \\&&( R_{Free-Bound}^{anti-parallel})^{2}+( R_{Bound-Bound}^{parallel})^{2}+( R_{Bound-Bound}^{anti-parallel})^{2}+\nonumber \\&& (R_{Free-Free}^{parallel} R_{Free-Free}^{anti-parallel})\partial^{2}(R_{Free-Free}^{parallel}+ R_{Free-Free}^{anti-parallel} )+\nonumber \\&&(R_{Free-Bound}^{parallel} R_{Free-Bound}^{anti-parallel})\partial^{2}(R_{Free-Bound}^{parallel}+ R_{Free-Bound}^{anti-parallel} )+\nonumber \\&&(R_{Bound-Bound}^{parallel} R_{Bound-Bound}^{anti-parallel})\partial^{2}(R_{Bound-Bound}^{parallel}+ R_{Bound-Bound}^{anti-parallel} )\label{wj8}
\end{eqnarray}

where the index $Bound-Bound$ denotes the interaction between  two bound electrons, $Free-Bound$ refers to the interaction of one free electron with one bound electron and $Free-Free$ denotes the interaction between two free electrons. Free electrons are electrons that aren't bound to specific atoms, while bound electrons are electrons that are bound to specific atoms. Substituting equations (\ref{wj2},\ref{wj4},\ref{wj7},\ref{wj8}) in equation (\ref{spm29}), putting $N=4$ and assuming that all corners are the same, we obtain the following result for one base \cite{R7,R8}:

 \begin{eqnarray}
&& E_{system}=\int d^{4}y \times\nonumber \\&&
\Biggl[\sqrt{-g}\Bigl(\Big(-(1-(\frac{1}{ \tilde{a}_{\alpha_{n'}}^{2}})^{2})
[( R_{Free-Free}^{parallel})^{2}+( R_{Free-Free}^{anti-parallel})^{2}+( R_{Free-Bound}^{parallel})^{2}+\nonumber \\&&( R_{Free-Bound}^{anti-parallel})^{2}+( R_{Bound-Bound}^{parallel})^{2}+( R_{Bound-Bound}^{anti-parallel})^{2}+\nonumber \\&& (R_{Free-Free}^{parallel} R_{Free-Free}^{anti-parallel})\partial^{2}(R_{Free-Free}^{parallel}+ R_{Free-Free}^{anti-parallel} )+\nonumber \\&&(R_{Free-Bound}^{parallel} R_{Free-Bound}^{anti-parallel})\partial^{2}(R_{Free-Bound}^{parallel}+ R_{Free-Bound}^{anti-parallel} )+\nonumber \\&&(R_{Bound-Bound}^{parallel} R_{Bound-Bound}^{anti-parallel})\partial^{2}(R_{Bound-Bound}^{parallel}+ R_{Bound-Bound}^{anti-parallel} )]+\nonumber\\&&
(\frac{1}{ \tilde{a}_{\alpha_{n'}}^{2}})^{2}\lambda^{2}\delta_{\rho_{1}\sigma_{1}}^{\mu_{1}\nu_{1}}
((R^{anti-parallel,\rho_{1}\sigma_{1}}_{Free-Free,\mu_{1}\nu_{1}}+R^{anti-parallel,\rho_{1}\sigma_{1}}_{Bound-Bound,\mu_{1}\nu_{1}}+R^{anti-parallel,\rho_{1}\sigma_{1}}_{Free-Bound,\mu_{1}\nu_{1}})-\nonumber\\&&(R^{parallel,\rho_{1}\sigma_{1}}_{Free-Free,\mu_{1}\nu_{1}}+R^{parallel,\rho_{1}\sigma_{1}}_{Bound-Bound,\mu_{1}\nu_{1}}+R^{parallel,\rho_{1}\sigma_{1}}_{Free-Bound,\mu_{1}\nu_{1}}))  \Bigr)\Big)\label{wj10}
\end{eqnarray}

Above equation shows that the energy of bases in each DNA depends on the curvature, which is created by the interaction of two bound electrons, a free and a bound electron and also two free electrons.  For the small size of a DNA molecule, this curvature is very strong and is a signature of the communication of bases with very distant bases in other DNA molecules.

\section{Modeling the electrical circuit of a DNA }\label{o1}

In this section, we show that the structure of a DNA molecule is very similar to the structure of  a radio receiver.  A simple radio receiver contains an antenna, an inductor or a coil, a capacitor, a detector or a diode  and a headphone (see Figure 6). A diode is a two-terminal electronic component that conducts primarily in one direction; it has low (ideally zero) resistance to the current in one direction, and high (ideally infinite) resistance in the other direction.   A capacitor stores energy in the electric field (E) between its plates. The amount of this energy depends on the voltage between its plates. An inductor stores energy in its magnetic field (B). The amount of this energy depends on the current that passes through it. The inductor and the capacitor together form a tuned or resonant circuit. This circuit can act as an electrical resonator, which stores energy. This energy is oscillating at the circuit's resonant frequency.  The resonant frequency of this  circuit is $f_{0}=\frac{1}{2\pi \sqrt{LC}}$ where $L$ is the inductance and $C$ is the capacitance. At this frequency,  the energy oscillates back and forth between the capacitor and the inductor.

\begin{figure*}[thbp]
	\begin{center}
		\begin{tabular}{rl}
			\includegraphics[width=12cm]{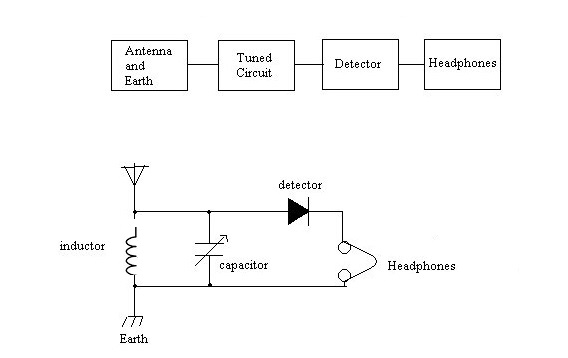}
		\end{tabular}
	\end{center}
	\caption{ The circuit of a simple radio receiver \cite{R1} }
\end{figure*}

We now compare the structure of a DNA molecule with the circuit of a radio receiver. To this aim, we begin with the structure of base pairs in DNA. It is clear that each A (Adenine)-T(Thymine) base pair contains  two electric dipoles between adenine and thymine, which have opposite directions with respect to each other. Also, each G (Guanine)-C (Cytosine) pair has three electric dipoles between the guanine and cytosine bases, and the direction of one of them is opposite with respect to the direction of the two other dipoles \cite{R2}.  The electric field of this  dipole is:

\begin{eqnarray}
&& d\sim 2\times 10^{-9} \quad q=\delta \times 1.6 \times 10^{-19} \quad r=\frac{d}{2} \quad \delta \approx 10^{-3} \nonumber\\&& E\sim \frac{\delta qd}{4\pi\varepsilon r^{3}}\sim 1.152 \times 10^{8} \label{D1}
\end{eqnarray}

where $d$ is the separation distance between positive and negative charges within a base pair, $q$ is the electric charge and $r$ is the place of one point between two bases.  The base pair has a length of 1--3 nanometers \cite{R3}. This results in a large amount of  energy stored in the electric field of dipoles between two bases  and thus the dipoles act like a capacitor. In AT base pairs, two capacitors act approximately in opposition to each other and the total electric field is zero (see Figure 7). However, in a GC base pair, one capacitor acts in opposition to two capacitors and the total electric field is not zero (see Figure 8). On the other hand, these strong electric fields conduct electrons in one direction and prevent their motion in the other direction and thus sometimes, base pairs act as a diode.

\begin{figure*}[thbp]
	\begin{center}
		\begin{tabular}{rl}
			\includegraphics[width=12cm]{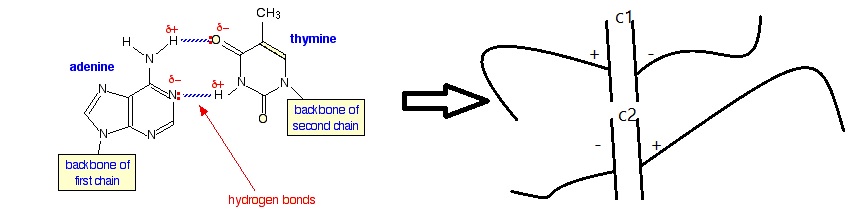}
		\end{tabular}
	\end{center}
	\caption{ The A-T base pair contains  two capacitors that act in opposition to each other.  }
\end{figure*}

\begin{figure*}[thbp]
	\begin{center}
		\begin{tabular}{rl}
			\includegraphics[width=12cm]{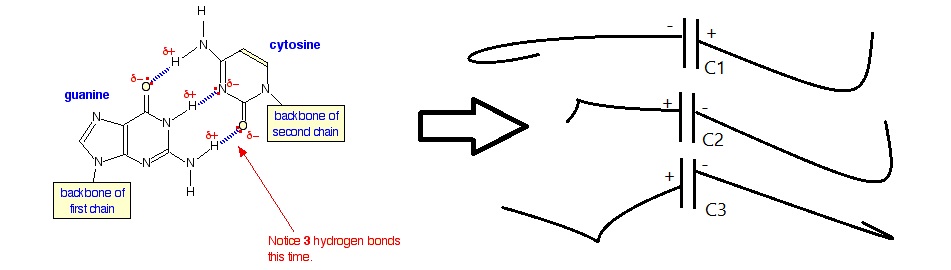}
		\end{tabular}
	\end{center}
	\caption{ The G-C base pair contains  three capacitors that act in opposition to each other.  }
\end{figure*}

Each DNA molecule  contains two biopolymer strands, which are coiled around each other to build a double helix. The nitrogenous bases of the two separate  strands are bounded together, according to base pairing rules (A with T, and C with G), with hydrogen bounds to make a double-stranded DNA molecule. Each base pair contains two or three electric dipoles, which produce an strong electric field along the coil. The electric fields of all bases are summed with each other and create a pure electric field along the coil. This electric field induces a force on electrons and creates a current along the coil of DNA. This leads to the emergence of an inductor in a DNA molecule (see Figure 9).

\begin{figure*}[thbp]
	\begin{center}
		\begin{tabular}{rl}
			\includegraphics[width=12cm]{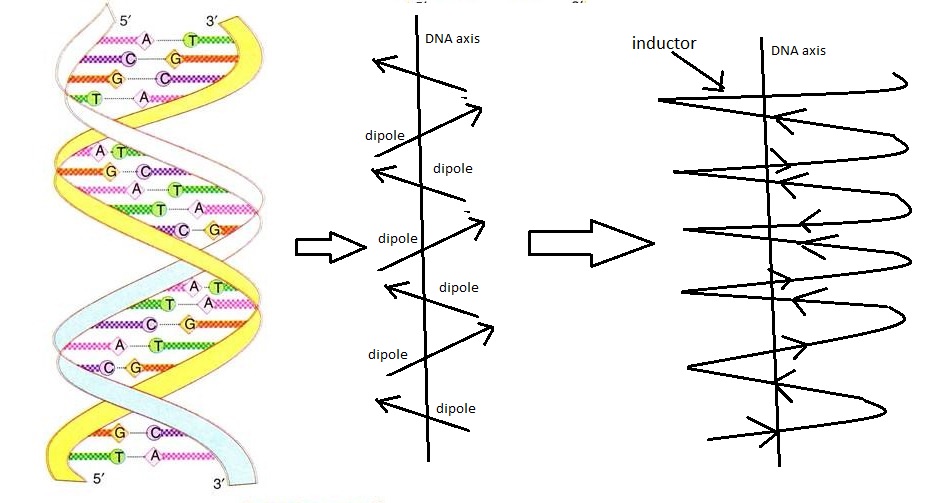}
		\end{tabular}
	\end{center}
	\caption{ Dipoles between bases in a DNA molecule form an inductor (coil).  }
\end{figure*}

The length of a DNA  is much greater than the size of the nucleus and for this reason, the DNA has to be packed tightly in a chromosome. The degree to which DNA is compressed is called its packing ratio  \cite{R4,R5}. In a chromosome, this packing occurs during several stages. In the first stage of packing, DNA winds around a protein core to produce a ``bead-like'' structure termed a nucleosome. This results in a packing ratio of about 6. The  second stage of packing is the coiling of beads in a helical structure called the 30 nm fiber that is found in all types of chromosomes. This structure increases the packing ratio to about 40. The final packaging happens when the fiber is organized in loops, scaffolds and domains that give a final packing ratio of about 1000 to 10000 in different types of chromosomes. During these stages, four types of inductor or coil appear. One coil emerges because of the electric dipoles in the structure of a DNA. A second type is produced by coiling DNA bases around the histone in a nucleosome (see Figure 10). A third type is created by the formation of loops in a chromatin fiber (see Figure 11) and finally, a fourth type of inductor emerges along the supercoil within a chromosome (see Figure 12).  Each of these inductor types produces one type of magnetic field and plays a main role in a resonant circuit.

\begin{figure*}[thbp]
	\begin{center}
		\begin{tabular}{rl}
			\includegraphics[width=12cm]{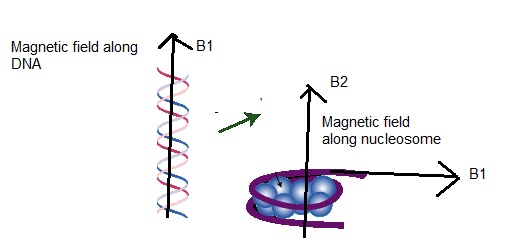}
		\end{tabular}
	\end{center}
	\caption{ Formation of an inductor or a coil along the nucleosome with two magnetic fields. }
\end{figure*}

\begin{figure*}[thbp]
	\begin{center}
		\begin{tabular}{rl}
			\includegraphics[width=12cm]{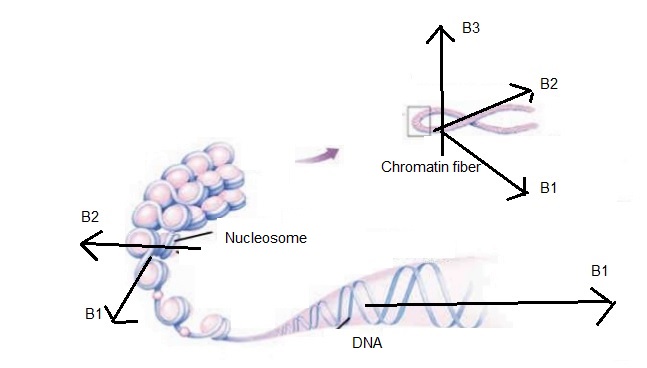}
		\end{tabular}
	\end{center}
	\caption{ Formation of an inductor or a coil along the chromatin fiber with three magnetic fields.}
\end{figure*}

\begin{figure*}[thbp]
	\begin{center}
		\begin{tabular}{rl}
			\includegraphics[width=12cm]{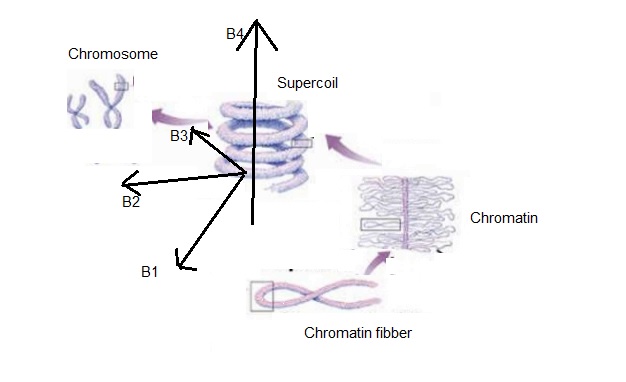}
		\end{tabular}
	\end{center}
	\caption{ Formation of an inductor or a coil along the supercoil with four magnetic fields.}
\end{figure*}

A chromosome has different parts, like a telomere, a satellite, a centromere and arms \cite{R6}. A telomere is the region of DNA at the end of the linear eukaryotic chromosome that is required for the replication and stability of the chromosome. This region is one of the main parts of a chromosome that has a direct effect on sending and receiving electromagnetic waves. For a human, the sequence of nucleotides in telomeres is TTAGGG, with the complementary DNA strand being AATCCC, with a single-stranded TTAGGG overhang. In a telomere, each  guanine has at least two positive and one negative induced charges. As we previously considered, a dipole can form between positive and negative charges and a capacitor emerges. Also, extra positive induced charges may have the role of free charges which move along the antenna (see Figure 13). If two guanine bases join each other, two extra positive charges lead to the emergence of electrical resistance between two capacitors (see Figure 14). In a  thymine base, there is only one positive and one negative induced charge. These  charges produce strong electric fields in one direction and induce a force to charges to move in this direction and prevent their motion in the opposite direction, just like a diode (see Figure 15). If we put electric devices that are similar to the bases of a telomore near each other, we can build a part of radio circuit like figure 16.

\begin{figure*}[thbp]
	\begin{center}
		\begin{tabular}{rl}
			\includegraphics[width=12cm]{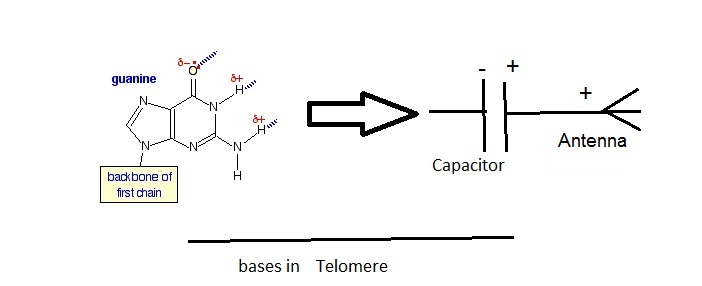}
		\end{tabular}
	\end{center}
	\caption{ Guanine in a telomere may play the role of capacitor + antenna in a circuit. }
\end{figure*}

\begin{figure*}[thbp]
	\begin{center}
		\begin{tabular}{rl}
			\includegraphics[width=16cm]{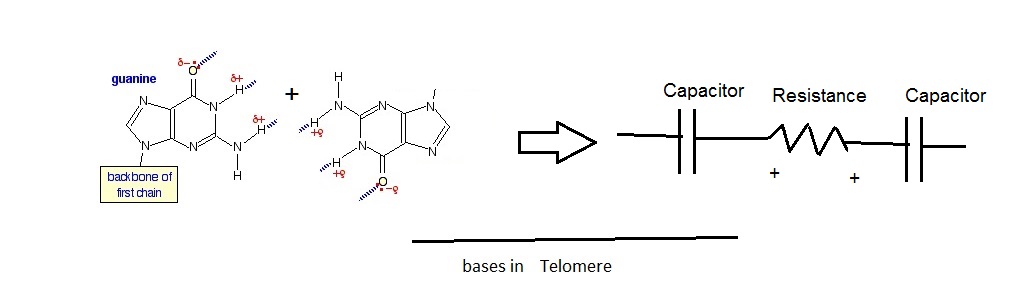}
		\end{tabular}
	\end{center}
	\caption{ Two guanine bases in a telomere may play the role of capacitor  + resistance + capacitor in a circuit. }
\end{figure*}

\begin{figure*}[thbp]
	\begin{center}
		\begin{tabular}{rl}
			\includegraphics[width=10cm]{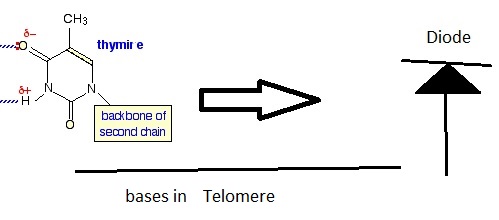}
		\end{tabular}
	\end{center}
	\caption{ Thymine in a telomere may play the role of a diode in a circuit. }
\end{figure*}

\begin{figure*}[thbp]
	\begin{center}
		\begin{tabular}{rl}
			\includegraphics[width=16cm]{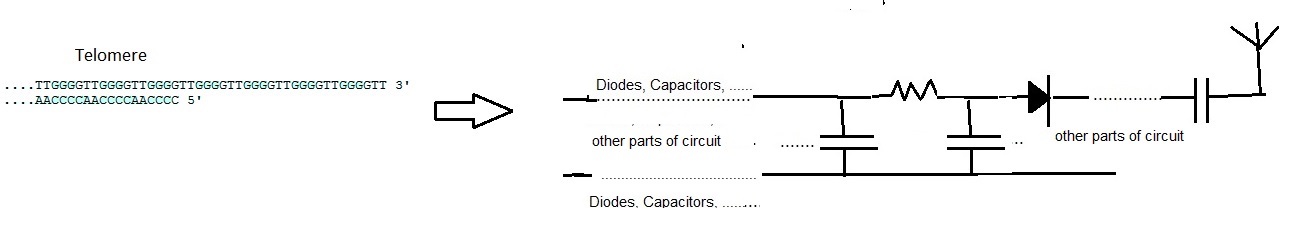}
		\end{tabular}
	\end{center}
	\caption{ Modeling the electrical circuit of a telomere.}
\end{figure*}

If we replace bases with electrical devices like inductors and capacitors, we can construct at least four tuned  or  resonant circuits.  Thus, each DNA has at least four  resonant frequencies and many diodes, resistors and antennae (Figure 17 gives only an approximate representation of a DNA molecule when its bases are replaced with electric devices). This circuit is very similar to the circuit of an FM radio.  In each DNA, there are some genes that have been constructed from a special number of AT and CG-base pairs. However, in each cell,  all genes aren't active and thus, the number of genes and  base pairs that are playing the role in electrical circuit in each chromosome is different from  chromosomes in other cells. This causes the resonant frequencies of chromosomes in each cell to be different. This difference ensures that each chromosome receives  only special electromagnetic messages from the brain, neurons, other cells and proteins. Also, each chromosome can emit special electromagnetic signals to other cells and inform them of its own activity. This long distance communication offers an opportunity for all cells in a human body to be in unison with each other.

\begin{figure*}[thbp]
	\begin{center}
		\begin{tabular}{rl}
			\includegraphics[width=16cm]{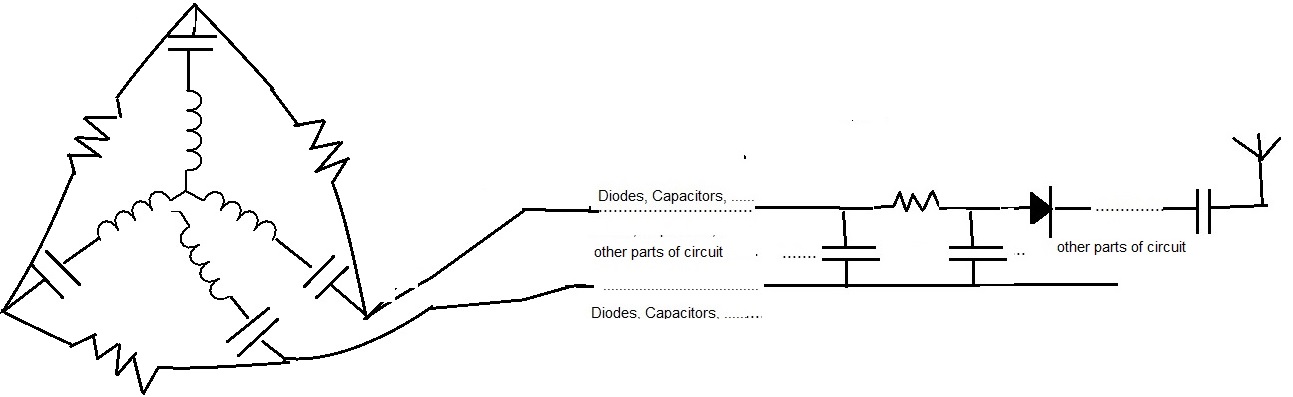}
		\end{tabular}
	\end{center}
	\caption{ Modeling a part of the  electrical circuit of DNA when bases are replaced by electric devices. }
\end{figure*}

\section{Damaged DNA, teleportation and cancer}\label{o2}

When a DNA molecule is damaged, its electric charges change and its radiated electromagnetic fields become different when compared to the signals of  a normal DNA molecule. For example, if one atom is added to one of bases in a DNA molecule, this base becomes a heptagonal molecule. Also, if a DNA molecule loses one of its atoms, the hexagonal shape of its base changes and a pentagonal molecule is created (see Figures 18, 19 and 20). Previously, for graphene,  it has been shown that the properties of a pentagonal molecule are different from hexagonal and heptagonal  ones. For example, a pentagonal molecule absorbs electrons, while a heptagonal molecule repels them \cite{R7, R8}.

\begin{figure*}[thbp]
	\begin{center}
		\begin{tabular}{rl}
			\includegraphics[width=5cm]{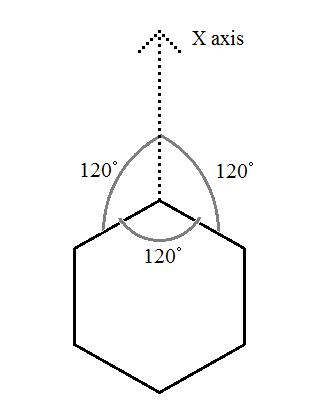}
		\end{tabular}
	\end{center}
	\caption{ The hexagonal shape of a molecule like molecules of bases in a DNA or molecules of a graphene. }
\end{figure*}

\begin{figure*}[thbp]
	\begin{center}
		\begin{tabular}{rl}
			\includegraphics[width=5cm]{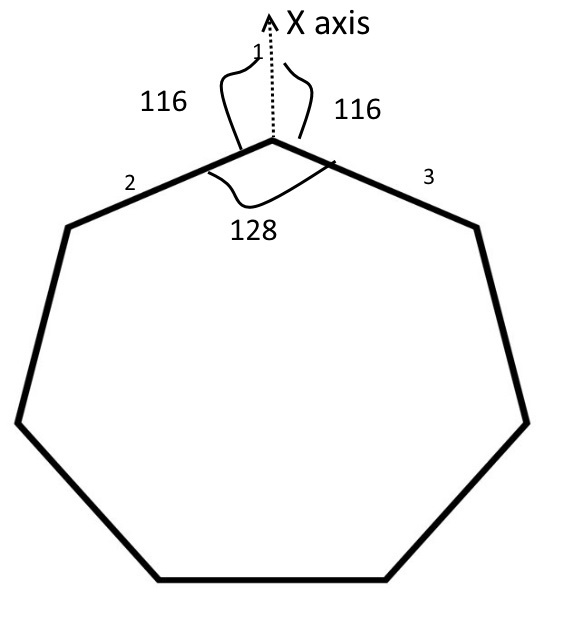}
		\end{tabular}
	\end{center}
	\caption{ The heptagonal shape of a molecule like molecules of bases in a DNA when one atom is added to them. }
\end{figure*}

\begin{figure*}[thbp]
	\begin{center}
		\begin{tabular}{rl}
			\includegraphics[width=5cm]{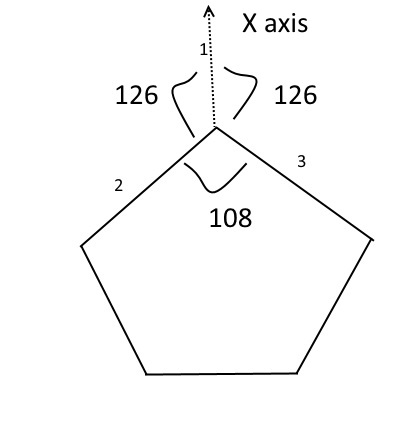}
		\end{tabular}
	\end{center}
	\caption{ The pentagonal shape of a molecule like molecules of bases in a DNA when lose one atom. }
\end{figure*}

In some recent research \cite{R7,R8}, the electric current for each of these molecules in graphene has been calculated. In these models, it has been asserted that the origin of gravitational waves and electromagnetic waves are the same and for this reason, all currents have been given in terms of curvature that is produced  between electrons \cite{R7,R8}:

\begin{eqnarray}
&&  J \approx \nonumber \\&&\sqrt{-g}\Bigl(\Big(-(1-m_{g}^{2})
[( R_{Free-Free}^{parallel})^{2}-( R_{Free-Free}^{anti-parallel})^{2}+( R_{Free-Bound}^{parallel})^{2}-\nonumber \\&&( R_{Free-Bound}^{anti-parallel})^{2}+( R_{Bound-Bound}^{parallel})^{2}-( R_{Bound-Bound}^{anti-parallel})^{2}+\nonumber \\&& (R_{Free-Free}^{parallel} R_{Free-Free}^{anti-parallel})\partial^{2}(R_{Free-Free}^{parallel}- R_{Free-Free}^{anti-parallel} )+\nonumber \\&&(R_{Free-Bound}^{parallel} R_{Free-Bound}^{anti-parallel})\partial^{2}(R_{Free-Bound}^{parallel}- R_{Free-Bound}^{anti-parallel} )+\nonumber \\&&(R_{Bound-Bound}^{parallel} R_{Bound-Bound}^{anti-parallel})\partial^{2}(R_{Bound-Bound}^{parallel}- R_{Bound-Bound}^{anti-parallel} )]+\nonumber\\&&
m_{g}^{2}\lambda^{2}\delta_{\rho_{1}\sigma_{1}}^{\mu_{1}\nu_{1}}
((R^{anti-parallel,\rho_{1}\sigma_{1}}_{Free-Free,\mu_{1}\nu_{1}}+R^{anti-parallel,\rho_{1}\sigma_{1}}_{Bound-Bound,\mu_{1}\nu_{1}}+R^{anti-parallel,\rho_{1}\sigma_{1}}_{Free-Bound,\mu_{1}\nu_{1}})+\nonumber\\&&(R^{parallel,\rho_{1}\sigma_{1}}_{Free-Free,\mu_{1}\nu_{1}}+R^{parallel,\rho_{1}\sigma_{1}}_{Bound-Bound,\mu_{1}\nu_{1}}+R^{parallel,\rho_{1}\sigma_{1}}_{Free-Bound,\mu_{1}\nu_{1}}))  \Bigr)\Big) \label{w1}
\end{eqnarray}

where $R_{Free-Free}^{parallel}$ is the curvature that is produced between two parallel free electrons in a DNA molecule, $R_{Free-Free}^{anti-parallel}$ is the curvature that is created between two anti-parallel free electrons, $R_{Free-Free}^{parallel}$ is the curvature that emerges between two parallel free electrons in a DNA molecule, $R_{Bound-Bound}^{parallel}$ is the curvature that is produced between two parallel bound electrons to atoms in a DNA molecule and $R_{Bound-Bound}^{anti-parallel}$ is the curvature that is created between two anti-parallel bound electrons to atoms in a DNA molecule. Also, $m_{g}$ is the graviton mass and $\lambda\sim \frac{1}{m_{g}}$. In additional, curvature is obtained from \cite{R7,R8}:

  \begin{eqnarray}
&& R_{Free/Bound-Free/Bound}^{anti-parallel}=l_{1}^{1-1}+l_{1}^{1-2}+l_{1}^{1-3}+\nonumber\\&&
(l'_{1})^{1-1}+(l'_{1})^{1-2}+(l'_{1})^{1-3}\nonumber\\&&\nonumber\\&& R_{Free/Bound-Free/Bound}^{parallel}=l_{2}^{1-1}+l_{2}^{1-2}+l_{2}^{1-3}-\nonumber\\&&
(l'_{2})^{1-1}-(l'_{2})^{1-2}-(l'_{2})^{1-3}\label{w2}
\end{eqnarray}

where $l_{1}$ is the coupling between two anti-parallel spins and $l_{2}$ is the coupling between parallel spins.

For  the corner of carbon-oxygen in a thymine, with hexagonal shape, we can measure these couplings in terms of the angle between axes in a molecule and electric charges of atoms (see Figure 15 and Figure 18):

\begin{eqnarray}
&& l_{1}^{1-1}\approx 2 q_{0}q_{1}\cos(0)=2 q_{0}q_{1} \quad l_{1}^{1-2}\approx q_{0}q_{2}\cos(120)=-\frac{1}{2}q_{0}q_{2} \quad l_{1}^{1-3}\approx q_{0}q_{3}\cos(240)=-\frac{1}{2}q_{0}q_{3}\nonumber\\&& (l'_{1})^{1-1}\approx \sin(0)=0 \quad (l'_{1})^{1-2}\approx q_{0}q_{2}\sin(120)=\frac{\sqrt{3}}{2}q_{0}q_{2} \quad (l'_{1})^{1-3}\approx q_{0}q_{3}\sin(240)=-\frac{\sqrt{3}}{2}q_{0}q_{3}\nonumber\\&& \nonumber\\&& l_{2}^{1-1}\approx q_{0}q_{1} \cos(60)=\frac{1}{2}q_{0}q_{1}  \quad l_{2}^{1-2}\approx q_{0}q_{2}\cos(180)=-q_{0}q_{2} \quad l_{2}^{1-3}\approx q_{0}q_{3}\cos(300)=\frac{1}{2}q_{0}q_{3}\nonumber\\&& (l'_{2})^{1-1}\approx q_{0}q_{1}\sin(60)=\frac{\sqrt{3}}{2}q_{0}q_{1}  \quad (l'_{2})^{1-2}\approx q_{0}q_{2}\sin(180)=0 \quad (l'_{2})^{1-3}\approx q_{0}q_{3}\sin(300)=-\frac{\sqrt{3}}{2}q_{0}q_{3}\label{w3}
\end{eqnarray}

where $q_{0}$ is the charge of the electron, $q_{1}$ is the charge of the oxygen atom, $q_{2}$ is the charge of the carbon atom,  $q_{3}$ is the charge of the nitrogen atom. We have assumed that half of electrons are parallel with each other and the other half are antiparallel with each other.  Substituting these values in equation (\ref{w2}), we can calculate the curvature for parallel and anti-parallel spins:

\begin{eqnarray}
&& R_{Free/Bound-Free/Bound}^{anti-parallel}=l_{1}^{1-1}+l_{1}^{1-2}+l_{1}^{1-3}+\nonumber\\&&
(l'_{1})^{1-1}+(l'_{1})^{1-2}+(l'_{1})^{1-3}=\nonumber\\&& 2 q_{0}q_{1} -\frac{1}{2}q_{0}q_{2}-\frac{1}{2}q_{0}q_{3} +\frac{\sqrt{3}}{2}q_{0}q_{2}-\frac{\sqrt{3}}{2}q_{0}q_{3}\nonumber\\&&\nonumber\\&& R_{Free/Bound-Free/Bound}^{parallel}=l_{2}^{1-1}+l_{2}^{1-2}+l_{2}^{1-3}-\nonumber\\&&
(l'_{2})^{1-1}-(l'_{2})^{1-2}-(l'_{2})^{1-3}=\nonumber\\&& \frac{1}{2}q_{0}q_{1}-q_{0}q_{2}+\frac{1}{2}q_{0}q_{3}-\frac{\sqrt{3}}{2}q_{0}q_{1}+\frac{\sqrt{3}}{2}q_{0}q_{3} \nonumber\\&& \label{w4}
\end{eqnarray}

By substituting the density of electrons for oxygen, nitrogen and carbon in equations (\ref{w1} and \ref{w4}), we can obtain the curvatures and current density:

\begin{eqnarray}
	&& q_{0}=\frac{1.6 \times 10^{-19}}{V_{Atom}} \quad q_{1}= \frac{8 q_{0}}{V_{Atom}} \quad  q_{2}= \frac{6 q_{0}}{V_{Atom}} \quad q_{3}=\frac{7 q_{0}}{V_{Atom}}\nonumber\\&& V_{Atom}=\frac{4\pi}{3} R^{3} \quad R\sim 10^{-10} \quad m_{g}\sim 0 \nonumber\\&& \nonumber\\&&\nonumber\\&& R_{Free/Bound-Free/Bound}^{anti-parallel}=88.576 \times 10^{22} \nonumber\\&&\nonumber\\&& R_{Free/Bound-Free/Bound}^{parallel}=5.22496 \times 10^{22} \nonumber\\&& \nonumber\\&&\nonumber\\&& J_{carbon-oxygen}\sim 7.818 \times 10^{48} \label{w5}
\end{eqnarray}

This result shows that the current density is positive and thus, free electrons are repelled by this corner of thymine. In parallel, we can calculate the current density in the corner of nitrogen-hydrogen:

\begin{eqnarray}
&&  J_{Nitrogen-Hydrogen}\sim -J_{carbon-oxygen}\sim -7.818 \times 10^{48} \label{w6}
\end{eqnarray}

 This equation indicates that the amount of current that goes outward of a thymine is equal to the amount of current that enters into  a thymine and the system is stable.

We now assume that a thymine base loses one of its atoms (For example a carbon atom) and assumes a pentagonal shape. For  carbon-oxygen corner in a thymine base, with hexagonal shape, we can repeat our calculations and obtain  couplings as a function of the angle between axes in a molecule and electric charges of atoms (see Figures 15 and 20):

 \begin{eqnarray}
 && l_{1}^{1-1}\approx 2q_{0}q_{1} \cos(0)=2q_{0}q_{1} \quad l_{1}^{1-2}\approx q_{0}q_{3}\cos(126)=-0.587q_{0}q_{3} \quad l_{1}^{1-3}\approx q_{0}q_{3}\cos(234)=-.587q_{0}q_{3}\nonumber\\&& (l'_{1})^{1-1}\approx \sin(0)=0 \quad (l'_{1})^{1-2}\approx q_{0}q_{3}\sin(126)=0.809q_{0}q_{3} \quad (l'_{1})^{1-3}\approx q_{0}q_{3}\sin(234)=-0.809q_{0}q_{3}\nonumber\\&& \nonumber\\&& l_{2}^{1-1}\approx 2q_{0}q_{1}\cos(60)=q_{0}q_{1}  \quad l_{2}^{1-2}\approx q_{0}q_{3}\cos(186)=-0.994q_{0}q_{3} \quad l_{2}^{1-3}\approx q_{0}q_{3}\cos(294)=0.406 q_{0}q_{3}\nonumber\\&& (l'_{2})^{1-1}\approx 2q_{0}q_{1}\sin(60)=1.732q_{0}q_{1}  \quad (l'_{2})^{1-2}\approx q_{0}q_{3}\sin(186)=-0.104q_{0}q_{3}\quad (l'_{2})^{1-3}\approx q_{0}q_{3}\sin(294)=-0.913q_{0}q_{3}\label{w7}
 \end{eqnarray}

Substituting these values in equation (\ref{w2}), we can calculate the curvature for parallel and anti-parallel spins:

\begin{eqnarray}
&& R_{Free/Bound-Free/Bound}^{anti-parallel}=l_{1}^{1-1}+l_{1}^{1-2}+l_{1}^{1-3}+\nonumber\\&&
(l'_{1})^{1-1}+(l'_{1})^{1-2}+(l'_{1})^{1-3}=\nonumber\\&& 2q_{0}q_{1} -0.587q_{0}q_{3}-.587q_{0}q_{3} \nonumber\\&& +0.809q_{0}q_{3} -0.809q_{0}q_{3} \nonumber\\&&\nonumber\\&& R_{Free/Bound-Free/Bound}^{parallel}=l_{2}^{1-1}+l_{2}^{1-2}+l_{2}^{1-3}-\nonumber\\&&
(l'_{2})^{1-1}-(l'_{2})^{1-2}-(l'_{2})^{1-3}=\nonumber\\&& q_{0}q_{1} -0.994q_{0}q_{3}+0.406 q_{0}q_{3} \nonumber\\&& - 1.732q_{0}q_{1} +0.104q_{0}q_{3}+0.913q_{0}q_{3} \label{w8}
\end{eqnarray}

Again, by substituting the density of electrons for oxygen, nitrogen and carbon atoms in equations (\ref{w1} and \ref{w8}), we can derive the curvature and current density:

\begin{eqnarray}
&& q_{0}=\frac{1.6 \times 10^{-19}}{V_{Atom}} \quad q_{1}= \frac{8 q_{0}}{V_{Atom}}  \quad q_{3}=\frac{7 q_{0}}{V_{Atom}}\nonumber\\&& V_{Atom}=\frac{4\pi}{3} R^{3} \quad R\sim 10^{-10} \quad m_{g}\sim 0 \nonumber\\&& \nonumber\\&&\nonumber\\&& R_{Free/Bound-Free/Bound}^{anti-parallel}=4.98048 \times 10^{22} \nonumber\\&&\nonumber\\&& R_{Free/Bound-Free/Bound}^{parallel}=-1.82692 \times 10^{22} \nonumber\\&& \nonumber\\&&\nonumber\\&& J_{carbon-oxygen}\sim 21.4747833 \times 10^{44} \label{w5}
\end{eqnarray}

This current density is much smaller than the current density that is produced by this corner (carbon-oxygen) in a normal DNA , however the neighboring atoms at other corners (nitrogen-hydrogen) aren't changed and thus, they produce the previous current density:

\begin{eqnarray}
&&  J_{Nitrogen-Hydrogen}\sim  -7.818 \times 10^{48}\nonumber\\&& \nonumber\\&&\nonumber\\&& J_{induced}\sim J_{Nitrogen-Hydrogen}-J_{carbon-oxygen}\sim 7.818 \times 10^{48}-21.4747833 \times 10^{44}\neq 0 \label{w6}
\end{eqnarray}

This extra current density produces many problems not only for this DNA molecule and the cell that contains it but also for other DNAs. In fact, a change in the current  of this DNA occurs and some extra signals are produced. These signals are received by inductors in other DNA molecules and an extra current density is created that leads to the destruction of other DNA molecules (see Figure 21). This event may be a reason for the production and duplication of cancer cells.

\begin{figure*}[thbp]
	\begin{center}
		\begin{tabular}{rl}
			\includegraphics[width=5cm]{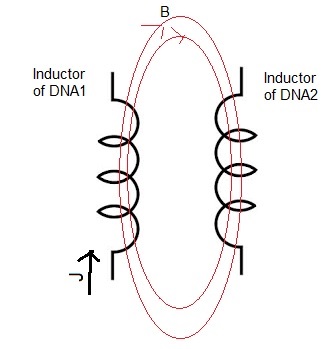}
		\end{tabular}
	\end{center}
	\caption{ The emitted signals by the inductor of one DNA can be detected by the inductor in other DNA. }
\end{figure*}

Now, we can suggest a mechanism for preventing of the progression of some diseases like cancer. If we add a third inductor to this system with properties that are the same as those of inductors of DNA and produce a current  in the opposite direction with respect to the current of the damaged DNA, an extra magnetic field emerges. This field cancels the effects of the magnetic field that is induced by the damaged DNA in normal DNA and prevents its destruction (see Figure 22). However, this is not a very simple method. To accomplish this, we need to know the exact form of the circuit in a DNA and design an exact circuit for communication with it.

\begin{figure*}[thbp]
	\begin{center}
		\begin{tabular}{rl}
			\includegraphics[width=5cm]{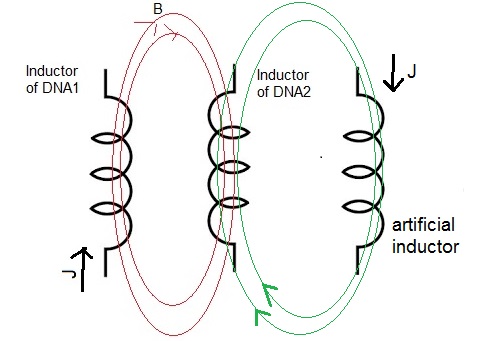}
		\end{tabular}
	\end{center}
	\caption{ The emitted signals by the inductor of a damaged DNA can be canceled by the artificial inductor . }
\end{figure*}

\section{The  difference between radiated waves from DNA molecules in men and women}\label{o3}

First, we design a simple circuit that measures radiated waves by DNA molecules of a man and a woman.  This circuit contains a ferromagnet  like an iron core, with two antiparallel magnetic fields are applied to it. These magnetic fields cause two regions to appear, with the spin of electrons in each of them parallel with each other and antiparallel with the spin of electrons in the other region (see Figure 23). This ferromagnet  is connected to a wire, resistance and ammeter. Our ammeter can measure voltages, too. Under normal conditions, electrons with upper spin in one region are attracted by electrons with lower spin in another region and the total current in the circuit becomes zero. If a single man comes close to this system, a negative current will be measured by the ammater, whereas the proximity of a single woman to it induces a positive current.

\begin{figure*}[thbp]
	\begin{center}
		\begin{tabular}{rl}
			\includegraphics[width=8cm]{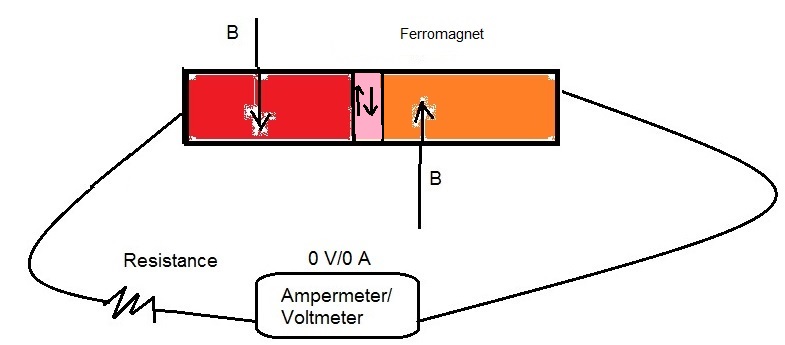}
		\end{tabular}
	\end{center}
	\caption{ The circuit of a simple circuit for the measurement of  the difference between radiated waves from DNAs in men and women  }
\end{figure*}

To discover the origin of the difference between radiated signals from man and women, we consider the difference in their chromosomes. Each human cell contains 23 pairs of chromosomes (22 pairs of autosomes or body chromosome(s) and one pair of sex chromosomes). This gives 46 chromosomes in total.  Autosomes emerge in pairs, whose members have the same form but differ from other pairs in a human cell. To have a stable system, it is necessary for the magnetic field in each member of one body chromosome or autosome to be equal in magnitude but opposite in direction to the magnetic field in the other member, such that they form a pair  without any magnetic field (see Figure 24).

\begin{figure*}[thbp]
	\begin{center}
		\begin{tabular}{rl}
			\includegraphics[width=5cm]{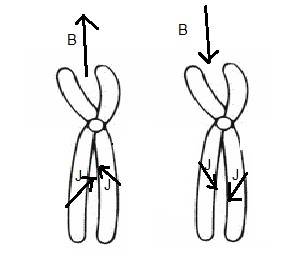}
		\end{tabular}
	\end{center}
	\caption{ The magnetic field of each member in an autosome pair is in the opposite direction to the magnetic field of the other member. }
\end{figure*}

Each cell contains one pair of sex chromosomes, one X and one Y chromosome in men and  two X chromosomes in women. One of X chromosomes in a man's cell is inactive. Also, the Y chromosome in a woman's cell with 59 million base pairs of DNA is shorter of the X chromosome in a man's cell with 155 million base pairs. These differences lead to the breakdown of stability in pairs and the radiation of two different signals. In the XX pair of a woman's cell, the magnetic field of one of chromosomes is turned off and thus, this pair radiates a magnetic field (see Figure 25). In the XY pair of a man's cell, the magnetic field of the X chromosome cannot be canceled by the magnetic field of the Y chromosome and a magnetic field is created in a direction that is opposite to that of the XX  pair in a woman's cell (see Figure 26).

\begin{figure*}[thbp]
	\begin{center}
		\begin{tabular}{rl}
			\includegraphics[width=8cm]{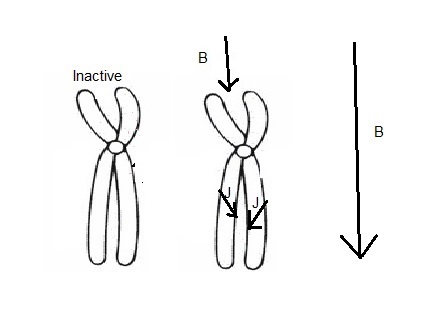}
		\end{tabular}
	\end{center}
	\caption{ In the XX pair of a woman's cell, the magnetic field of one of chromosomes is turned off and thus, this pair radiates a magnetic field. }
\end{figure*}

\begin{figure*}[thbp]
	\begin{center}
		\begin{tabular}{rl}
			\includegraphics[width=8cm]{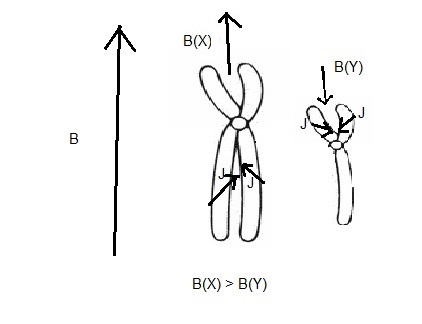}
		\end{tabular}
	\end{center}
	\caption{In the XY pair of a man's cell, the magnetic field of the X chromosome cannot be canceled by the magnetic field of the Y chromosome and a magnetic field is created in a direction opposite to that of the XX  pair in a woman's cell. }
\end{figure*}

When a DNA molecule is damaged, an extra current appears along its inductor, which produces an extra electromagnetic field. This extra wave induces an extra current in other DNA molecules and leads to their destruction and disease progression. This event can be seen in some diseases like cancer. If we put a damaged DNA of a male or a female near the damaged DNA of a female or a male, their radiated waves cancel the effect of each other and disease progression is stopped (see Figure 27).

\begin{figure*}[thbp]
	\begin{center}
		\begin{tabular}{rl}
			\includegraphics[width=8cm]{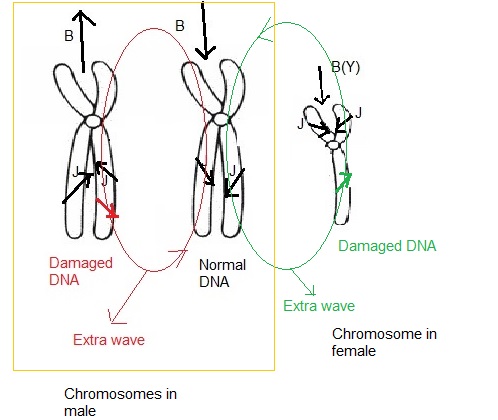}
		\end{tabular}
	\end{center}
	\caption{To remove extra signals of a damaged DNA in males, we can use of damaged DNA in females }
\end{figure*}

Another model to cure cancer is that we  send some electromagnetic signals and force cancer cells   to return to their initial stem cells. Then, by controlling their differentiating,  we can produce some new types of non-cancer cells.

Another application of the difference between the  radiated electromagnetic signals of men and women is the determination of the gender of a fetus in pregnant women. When the fetus is female, its radiated waves have the same sign and oscillation as the radiated waves of the mother and amplify her signals, while if the fetus is male, its radiated waves have the opposite sign and oscillation compared to those of the mother and thus they destroy her signal.  This method helps us determine the gender of a fetus  much sooner than normal methods like sonography. Also, this mechanism is more harmless when compared to sonography.

Another interesting result that comes out of these calculations is the difference in the electromagnetic signals of sperms and eggs. In fact, sperms that produce males have different signals respect to the sperms that produce females.  This  may help us to determine the gender of a fetus which is produced by the fusion of  an special sperm with an special egg.

Dad cell and Mam cell are stable and constructed from chromosomes that their radiated waves cancel the effect of each other. By dividing Dad cell, two sperms are emerged that their radiated waves have opposite signs. Also, by dividing Mam cell, two eggs are produced that their waves have opposite signs. By joining an sperm and an egg with opposite waves, a diploid cell is created. It is clear that the created cell isn't stable and may emit waves with positive sign or negative sign. To determine the gender of a fetus, we can put two types of sperms near it. Sperms which radiates waves with the same sign of fetus, repel it and becomes far from it and sperms that radiates waves with opposite sign of fetus, become close to it. This method is more easier than normal mechanisms like sonography.

\begin{figure*}[thbp]
	\begin{center}
		\begin{tabular}{rl}
			\includegraphics[width=10cm]{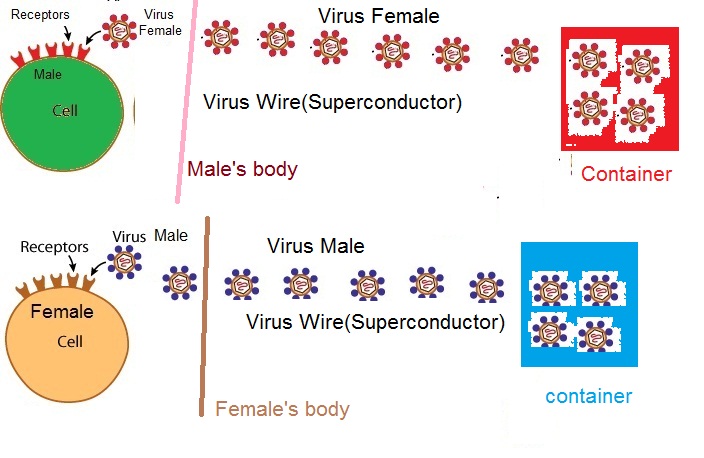}
		\end{tabular}
	\end{center}
	\caption{The Female Virus Wire (Superconductor) for Males and the Male Virus Wire (Superconductor) for Females. }
\end{figure*}

\begin{figure*}[thbp]
	\begin{center}
		\begin{tabular}{rl}
			\includegraphics[width=10cm]{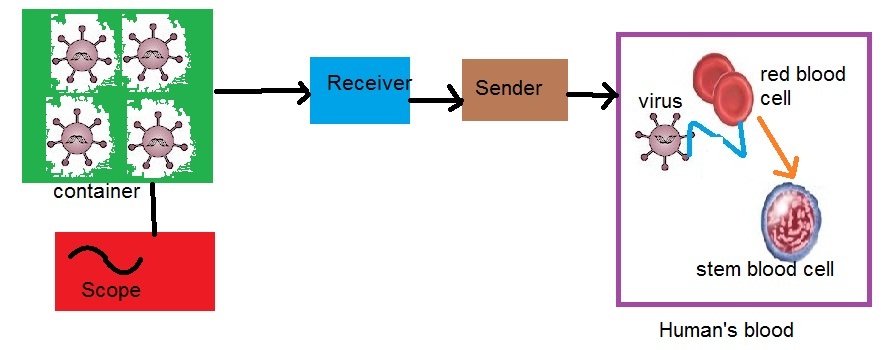}
		\end{tabular}
	\end{center}
	\caption{The Virus Medical Imaging Technique: Viruses in blood take signals of viruses outside the blood in VMI machine and controll the process of differentiation. }
\end{figure*}

\begin{figure*}[thbp]
	\begin{center}
		\begin{tabular}{rl}
			\includegraphics[width=8cm]{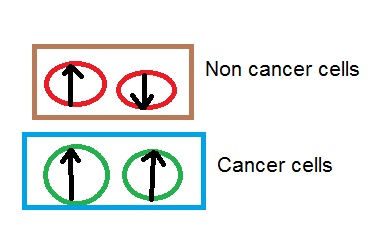}
		\end{tabular}
	\end{center}
	\caption{The radiated waves of two cells in neighbour is in reverse and total wave is zero. For cancers, total wave isnot zero. }
\end{figure*}

We can generalize this discussion to viruses. We can show that viruses which are attracted by cells of a female have different signals respect to viruses which are absorbed by cells of a male. Although, type of these viruses are the same, however, packing of DNA or the location of chromosomes is different which leads to the radiation of two different signals. In fact, some of viruses are male and some others are female. On the other hand, the reason for the emergence of some cancers are viruses. Thus, we can remove the effects of viruses which lead to the cancer with viruses of opposite gender.  Another interesting application of viruses is the  constructing a wire  that transmit information very fast like wormhole in cosmology or superconductor in nano. We can call them virus wormhole or virus superconductor (See Figure 28). These wires consist of different type of viruses and penetrate in human's body.

Another interesting point about viruses is their application as the receiver or sender of electromagnetic waves. We can use of some viruses as the wireless devises for communications with DNAs in various cells. Also, they can help us to communicate with specialized cells and force them to return to initial stem cell or even initial sperm and egg.

Above results help us to introduce two types of medical imaging techniques:1.Sperm Medical Imaging (SMI)2.Virus Medical Imaging (VMI). In Sperm Medical Imaging technique, we use of the communications between sperms and cells. In this method, by decoding the radiated signals of sperms and considering  their evolutions, we can imagine the events interior of a human's body.  In Virus Medical Imaging, we use of the communications between Viruses and cells interior of human's body. In this model, by considering the evolutions of viruses out of a body, we can analyse the events interior of a human's body. 

To construct a device that communicate with DNAs of cells interior of human's body, we use of the MRI machin which it's protons are replaced by sperms or viruses. In Sperm/Virus Medical Imaging, signals from sperms/Viruses are received by some electronic devices or some special antennas and then applied to the patient. This leads to the occurrence of  a communication  between sperms/Viruses and cells. The DNAs in cells interior of human's body or tissue response to sperms and send  radio frequency waves, which are taken by a radio receiver. These signals may be decoded and the exact information about the evolutions of cells interior of a body can be obtained (See Figure 29).

In human's body, the radiated waves of  cells cancel the effect of each other and total radiation of body is approximately zero (See figure 30). If these waves couldn't remove the effect of each other, some extra information is transmitted between cells and cancer may be emerged. In these conditions, we can communicate with cancer cells via electromagnetic cells and force them to transform to stem cell. By this the problem of cancer may be solved.

One of interesting subject in biology is the ability of DNA for saving information. We can show that water has the same ability and we can define a coding for it like the genetic coding of DNA. Also, electromagnetic waves have an special type of coding and act similar to DNA or protein. When electromagnetic waves achieve to molecules of water, give their information to water and take it's information. The process is very the same of transcription and translation in biological systems. Sometimes, electromagnetic waves act like DNA and water acts like the protein or RNA and sometimes, reversely, water acts like the DNA and electromagnetic waves act like the protein.  On the other hand, water and electromagnetic waves have special genders. For example, molecules of water attract waves with opposite gender. 

In a DNA teleportation, two tubes or container of water become located near each other. The first tube emit some signals that contain some packages. These packages carry the information about structures which are formed in initial water. When these packages achieve to second tube of water, open and give their information to second water (See figure 31). We can write:

\begin{eqnarray}
	&& H^{EM}=  H^{tot}=  \Sigma_{j=3}^{N-3}\Sigma_{i=1}^{N}P_{ij}H_{j,i}^{tot} = \nonumber\\ && \Sigma_{j=3}^{N-3}\Sigma_{i=1}^{N}EM_{j,i}^{tot} 
	\label{u4}
\end{eqnarray}

where $EM_{j,i}^{tot}$ is the package of information in an electromagnetic wave which produces j number of i-gonal structures with the Humiltonian of $H_{j,i}^{tot}$ and probability of $P_{ij}$. Molecules of water have trigonal shapes, however in a tube, they join each other and form other structures like hexagonal shapes.  This type of writing for the Humiltonian of electromagnetic waves helps us to introduce a new information codding for waves.
 
\begin{figure*}[thbp]
	\begin{center}
		\begin{tabular}{rl}
			\includegraphics[width=8cm]{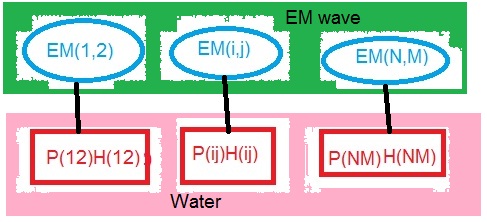}
		\end{tabular}
	\end{center}
	\caption{ Coding of water and electromagnetic waves }
\end{figure*}

If we put DNA in initial tube or container of water, this water emit some electromagnetic waves that carry the information about the changes  produced by the  DNA in the structure of initial pure water . We can write:

\begin{eqnarray}
	&& H^{EM, DNA-Water}=  H^{tot, DNA-Water}=  \Sigma_{j=3}^{N-3}\Sigma_{i=1}^{N}\Sigma_{k=1}^{M}P_{ij,k}H_{j,i,k}^{tot, DNA-Water} = \nonumber\\ && \Sigma_{j=3}^{N-3}\Sigma_{i=1}^{N}\Sigma_{k=1}^{M}EM_{j,i,k}^{tot, DNA-Water} 
	\label{sdu4}
\end{eqnarray}

where $EM_{j,i,k}^{tot, DNA-Water} 
$ is the package of information in an electromagnetic wave which produces j number of i-gonal structures and k extra corners corresponded to k atoms of DNA. Also, $H_{j,i,k}^{tot, DNA-Water}$ is the Hamiltonian of these structures and $P_{ij,k}$ is the probability for producing them. When these packages of waves achieve to another pure water, form structures of initial water with some extra corners or defects which are the signature of the existence of DNA. For example, assume that a tube or container of water contains DNA molecules. These molecules change the hexagonal and pentagonal structures of water. Signals which are produced by this water contain packages which carry the information of structures of initial water. When these packages achieve to second tube of pure water, give their information to water, change structures and produce the same of structures of initial water. We can detect the structures in initial water by using the PCR or other methods. 

We can use of the gender of water in medical imaging. First, we separate two waters one with the gender of male which interacts with  men and another with the gender of female which interacts with women. If  a woman eat water wit the gender of female,
 all of it repel with her body. For a pregnant women with a fetus of male, some of this water are absorbed, while for a fetus of female, all of this water go out. This method is completely harmless respect to other methods like the Sonography. Another interesting pint is separating waves by using water. Water with the gender of male attract waves with the gender of female and water with the gender of male, attract waves with the gender of female. Using this property, we can analyse the events interior of human's body by considering the exchanged waves between waters inside calls and outside of body.
 
 \begin{figure*}[thbp]
	\begin{center}
		\begin{tabular}{rl}
			\includegraphics[width=8cm]{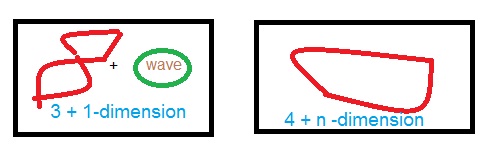}
		\end{tabular}
	\end{center}
	\caption{Waves (Topoisomerases) which  participate in the overwinding or underwinding of DNA in extra dimensions }
\end{figure*}

 In additional to electromagnetic waves, DNAs emit some new types of waves which have the shape of topoisomerases. Topoisomerases are enzymes that participate in the overwinding or underwinding of DNA. We generalise this concept to waves which  participate in the overwinding or underwinding of DNA in extra dimensions (See figure 32).  These wave topoisomerases have 4 + n -dimensional shapes. This is because that DNA  is very compacted in 3 + 1 dimensional space-time. However if we see topology of DNA in extra dimension, we can open packings of it and obtain a simpler shape. In fact DNA acts like a  4 + n-dimensional coil which has been packed to 3 + 1-dimensions. Only, waves could interact with DNA, which could move in 4 +  n dimensions and see the un-packed structure of it in extra dimensions. These waves couldn't be classified as the electromagnetic or gravitational waves. However, by reducing dimensions to 4 or 3, they obtain some properties similar to electromagnetic waves. 

 When, a DNA is packed, some extra fields are emerged that play the role of extra photons in 4-dimensions. These fields change the action and Hamiltonian of DNAs and produce some new wormhole-like tunnels. For example, the interaction of two atoms can be written as:

\begin{eqnarray}
&& S_{atom-atom}=-T_{tri} \int d^{11}\sigma \sqrt{\eta^{ab}
	g_{MN}\partial_{a}X^{M}\partial_{b}X^{N} +2\pi
	l_{s}^{2}U(F)+ +2\pi
	l_{s}^{2}\bar{U}(\bar{F})}\nonumber\\&&
U=(\sum_{n=1}^{Z}\frac{1}{n!}(-\frac{F_{1}..F_{n}}{\beta^{2}}))
\nonumber\\&&
\bar{U}=(\sum_{n=1}^{\bar{z}}\frac{1}{n!}(-\frac{\bar{F}_{1}..\bar{F}_{n}}{\beta^{2}}))
\nonumber\\&& F=F_{\mu\nu}F^{\mu\nu}\quad
F_{\mu\nu}=\partial_{\mu}A_{\nu}-
\partial_{\nu}A_{\mu}\label{7}
\end{eqnarray}

where z is the number of paired electrons, $\bar{z}$ is the number of photons which are produced by the packings of DNA,  $g_{MN}$ is the background metric, $ X^{M}(\sigma^{a})$'s
are scalar fields which are constructed from paring two electrons with opposite spins, $\sigma^{a}$'s are
the manifold coordinates, $a, b = 0, 1, ..., 3$ are world-volume
indices of the manifold and $M,N=0, 1, ..., 11$ are eleven dimensional
spacetime indices. Also, $G$ is the nonlinear field
 and $A$ is the photon which exchanges between
atoms. Also, $\bar{A}$ is the photon which is emerged by te packings of DNA. We can obtain:

\begin{eqnarray}
&&H_{atom-atom}=4\pi T_{tri}\int d\sigma_{11}... d\sigma_{1}[(1 + Q_{z,atom-atom}) (1 + \bar{Q}_{\bar{z},atom-atom})]\times\nonumber\\ && \sqrt{1+\eta^{ab}g_{MN}g_{MN}\psi^{M,\uparrow}_{atom1}\partial_{a}\psi^{M,\downarrow}_{atom2}\psi^{N,\uparrow\uparrow}_{atom2}\partial_{b}\psi^{N,\downarrow}_{atom1}}\times\nonumber\\ && \sqrt{1+\eta^{ab}g_{MN}g_{MN}\bar{\psi}^{M,\uparrow}_{atom1}\partial_{a}\bar{\psi}^{M,\downarrow}_{atom2}\bar{\psi}^{N,\uparrow\uparrow}_{atom2}\partial_{b}\bar{\psi}^{N,\downarrow}_{atom1}}
\nonumber\\ && \nonumber\\ &&
\bar{Q}_{\bar{z},atom-atom}=\bar{Q}_{\bar{z}-1,atom-atom}\sqrt{1+\frac{k^{2}_{2}}{\bar{Q}_{\bar{z}-1,atom-atom}\theta^{4}_{\bar{z},atom-atom}}}....\sqrt{1+\frac{k^{2}_{2}}{\bar{Q}_{1,atom-atom}\theta^{4}_{1,atom-atom}}}
\nonumber\\
&& \bar{Q}_{1,atom-atom}= \sqrt{1+\frac{k^{2}_{1}}{\theta^{4}_{1,atom-atom}}}
 \nonumber\\ &&
Q_{z,atom-atom}=Q_{z-1,atom-atom}\sqrt{1+\frac{k^{2}_{2}}{Q_{z-1,atom-atom}\sigma^{4}_{z,atom-atom}}}....\sqrt{1+\frac{k^{2}_{2}}{Q_{1,atom-atom}\sigma^{4}_{1,atom-atom}}}
\nonumber\\
&& Q_{1,atom-atom}= \sqrt{1+\frac{k^{2}_{1}}{\sigma^{4}_{1,atom-atom}}}
\label{8}
\end{eqnarray}

where $\sigma$ is the separation distance between two atoms and $\theta$ is the angle of packings. Also, packing of DNA changes the couplings between fermions and produce some new couplings which are presented by $bar{\psi}$. Generalizing these calculations to all molecules of DNA and following the method in previous section, we can obtain:
 
\begin{eqnarray}
&&H_{DNA}=4\pi T_{tri}\int d\sigma_{11}... d\sigma_{1} Q_{DNA}E_{DNA}\nonumber\\ && Q_{DNA} =\Sigma_{M,N,X,Y=1}^{W} Q_{un-packed,DNA,N,M}\bar{Q}_{packed,DNA,X,Y}\nonumber\\ && E_{DNA} =\Sigma_{M,N,X,Y=1}^{W} E_{un-packed,DNA,N,M}\bar{E}_{packed,DNA,X,Y} \nonumber\\ &&   Q_{un-packed,DNA, M,N}= \Pi_{n=1}^{N}[\Sigma_{i,j=1}^{n} \big(1 + Q_{6,i-j} \big)] \Pi_{m=1}^{M}[\Sigma_{l,k=1}^{m} \big(1 + Q_{5,i-j} \big)]\times \nonumber\\ && \Pi_{n=1}^{N}[\Sigma_{i,j=1}^{n} \big(1 + Q_{6-5,i-j} \big)] \Pi_{m=1}^{M}[\Sigma_{l,k=1}^{m} \big(1 + Q_{5-6,i-j} \big)]\nonumber\\ &&E_{un-packed,DNA,M,N}= \Pi_{n=1}^{N}[\Sigma_{i,j=1}^{n} \big(1 + E_{6,i-j} \big)] \Pi_{m=1}^{M}[\Sigma_{l,k=1}^{m} \big(1 + E_{5,i-j} \big)]\times \nonumber\\&&  \Pi_{n=1}^{N}[\Sigma_{i,j=1}^{n} \big(1 + E_{6-5,i-j} \big)] \Pi_{m=1}^{M}[\Sigma_{l,k=1}^{m} \big(1 + E_{5-6,i-j} \big)] \nonumber\\ &&   \bar{Q}_{packed,DNA,X,Y}= \Pi_{n=1}^{N}[\Sigma_{i,j=1}^{n} \big(1 + \bar{Q}_{6,i-j} \big)] \Pi_{m=1}^{M}[\Sigma_{l,k=1}^{m} \big(1 + \bar{Q}_{5,i-j} \big)]\times \nonumber\\ && \Pi_{n=1}^{N}[\Sigma_{i,j=1}^{n} \big(1 + \bar{Q}_{6-5,i-j} \big)] \Pi_{m=1}^{M}[\Sigma_{l,k=1}^{m} \big(1 + \bar{Q}_{5-6,i-j} \big)]\nonumber\\ && \bar{E}_{packed,DNA, X,Y}= \Pi_{n=1}^{N}[\Sigma_{i,j=1}^{n} \big(1 +  \bar{E}_{6,i-j} \big)] \Pi_{m=1}^{M}[\Sigma_{l,k=1}^{m} \big(1 + \bar{E}_{5,i-j} \big)]\times \nonumber\\&&  \Pi_{n=1}^{N}[\Sigma_{i,j=1}^{n} \big(1 + \bar{E}_{6-5,i-j} \big)] \Pi_{m=1}^{M}[\Sigma_{l,k=1}^{m} \big(1 + \bar{E}_{5-6,i-j} \big)]
\label{9}
\end{eqnarray}

where $w = M + N + X + Y$ is the number of total fields which are emerged in a packed DNA. M, N are numbers of fields in un-packed DNA and X, Y are number of fields which are produced by packings of DNA.  This system is very complicated and waves couldn't read it's information easily (See figure 4). They should  open packings of DNA in extra dimensions and make it's topology simple. For this reason, it is needed that the waves give an appropriated energy to DNA, excited and open it's packings. In these conditions, total energy and topology system tend to a constant number. We can write: 

\begin{eqnarray}
&& 1= H_{DNA} + H_{topoisomerase-like wave} = \nonumber\\ && 4\pi T_{tri}\int d\sigma_{11}... d\sigma_{1} [Q_{DNA}E_{DNA} + Q_{topoisomerase}E_{topoisomerase}]= \nonumber\\ && \frac{4\pi^{w}}{V_{11-w}}\int d\sigma_{11}... d\sigma_{1}\Pi_{n=1}^{w} [\delta (\sigma_{n} + \theta_{n})] 
\label{10}
\end{eqnarray}

where we have replaced $4\pi T_{tri}$ with $\frac{4\pi^{w}}{V_{11-w}}$. Here, $V_{11-w}$ is the volume of space which is empty od DNA. To obtain this delta function, we can use of waves that number of packing fields in them is equal to number of un-packing fields of DNA and also, number of un-packing fields in them is equal to number of packings in DNA. To this aim, we put $X,Y$ instead of $M,N$ and also $M,N$ instead of $X,Y$ in Hamiltonian of DNA (equation (\ref{9})) and write:

\begin{eqnarray}
&&H_{topoisomerase-like wave}=4\pi T_{tri}\int d\sigma_{11}... d\sigma_{1} Q_{topoisomerase}E_{topoisomerase}\nonumber\\ && Q_{topoisomerase} =\Sigma_{M,N,X,Y=1}^{W} Q_{un-packed,topoisomerase,X,Y}\bar{Q}_{packed,topoisomerase,M,N}\nonumber\\ && E_{topoisomerase} =\Sigma_{M,N,X,Y=1}^{W} E_{un-packed,topoisomerase,X,Y}\bar{E}_{packed,DNA,M,N} \label{11}
\end{eqnarray}

Using equations (\ref{9} and \ref{11}), we can obtain the explicit form of delta function in equation (\ref{11}):

\begin{eqnarray}
&& \delta(\sigma_{n} -\theta_{n})= \frac{1}{\pi} \frac{[K_{\theta_{n}} + K_{\sigma_{n}}]^{2}}{[K_{\theta_{n}} + K_{\sigma_{n}}]^{2} + [\theta_{n} + \sigma_{n}]^{2}} \label{12}
\end{eqnarray}

Above results show that to exchange information between DNAs,  it is needed to some waves which their structures are similar to topoisomerases in biology. In these systems, number of packed manifolds is equal to number of un-packed manifolds of DNA and number of un-packed manifolds is equal to number of packed manifolds in DNA. By joining these waves to DNA, Hamiltonian and topology of system tends to a constant number. In these conditions, all information of DNA can be recovered and exchanged.

 \begin{figure*}[thbp]
	\begin{center}
		\begin{tabular}{rl}
			\includegraphics[width=8cm]{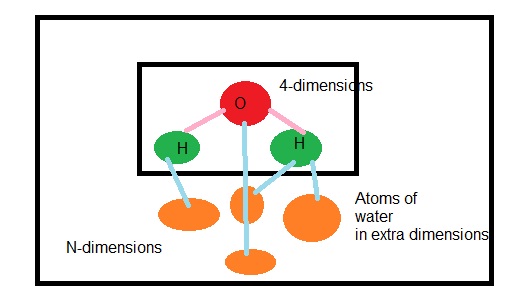}
		\end{tabular}
	\end{center}
	\caption{Extra atoms of water in extra dimensions. }
\end{figure*}

To understand the reason for water memory,we should go to extra dimensions and consider the real structure of water (See figure 33). In fact, water has a coding structure that only part of it can be seen in 4-dimensions. Also,waves have coding structures that only part of their structures can be detected in 4-dimensions. However,by considering amount of memory of water and waves,we can achieve to their real structure in extra dimensions. Also,in extra dimensions,the differences between two genders of water can be cleared more and their structures could be different. Also, an observer in extra dimensions can see two types of gender for waves.Each water can interact with water of opposite gender. Reducing dimensions from 11 or more to 4, some of information about structures of water and waves is lost.

 \begin{figure*}[thbp]
	\begin{center}
		\begin{tabular}{rl}
			\includegraphics[width=8cm]{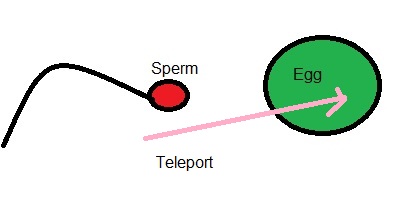}
		\end{tabular}
	\end{center}
	\caption{Teleportation of sperms into eggs via extra dimensions. }
\end{figure*}

One of examples of teleportation via waters is the teleportation of sperms into the  egg (See figure 34). When water of men which includes sperms become near to the water of women which includes egg, they become entangled. When two matters be very entangled, repel other object that be in opposed with this entanglement. For this reason, one of sperms is repelled by waters and transfer into eggs. Another cause of this teleportation is the entanglement between water inside the egg and water outside it. Sperms radiate some topoisomerse-like waves. These waves create some holes in egg. To fill these holes, sperm should be transferred to egg.

Another main subject is the magnetic fields of earth. It seems that magnetic fields of earth are more complicated of what they are seemed and have structures like topoisomerse-like waves. For this reason, they can contribute in teleportation. They produce some holes appropriated with objects that for filling them, objects should be transferred.

 \begin{figure*}[thbp]
	\begin{center}
		\begin{tabular}{rl}
			\includegraphics[width=8cm]{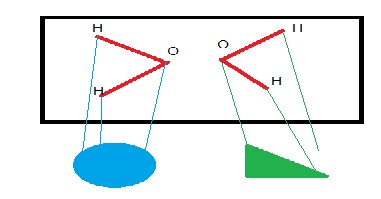}
		\end{tabular}
	\end{center}
	\caption{Two structures for water in extra dimensions:waters are the same in four dimensions and different in extra dimensions }
\end{figure*}

 \begin{figure*}[thbp]
	\begin{center}
		\begin{tabular}{rl}
			\includegraphics[width=8cm]{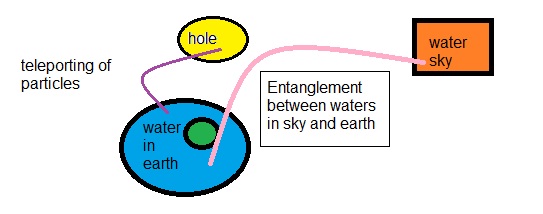}
		\end{tabular}
	\end{center}
	\caption{Teleportation of particles into holes in entangled sstems of waters of earth and sky. }
\end{figure*}

 \begin{figure*}[thbp]
	\begin{center}
		\begin{tabular}{rl}
			\includegraphics[width=8cm]{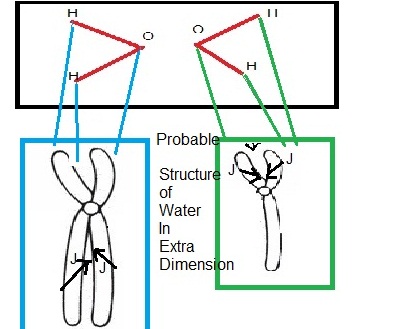}
		\end{tabular}
	\end{center}
	\caption{DNA-like structures of waters in extra dimensions. }
\end{figure*}

Another application of two structures for water in extra dimensions(See figure 35) and waves is radiating two types of  topoisomerase-like waves from earth and sun or other objects in the sky. The origin of these waves are unclear. About earth, it seems that the liquid in it's core produce these types of waves. These waves are needed for activity of cells and growing plants.  Also, these waves help in exchanging information between DNAs and contribute in transcription and translation. About radiated topoisomerase-like waves from sun, we guess that there are some waters in the sky that their waves mix with radiated waves of sun. Although,the structure of sun itself is yet questionable. The radiated waves from sun and earth create a medium that some points of it are holes and empty of fields. To fill these holes, particles could be teleported to them.  In fact,waters of earth and sky are entangled. Each particle enters into these waters, break the entanglement and produces some holes. To fill these holes, particles are teleported into these holes.From this point,we can use for transferring of objects and fields in future (See figure 36).  

Another question arises that what is the real structure of waters in extra dimensions (See figure 37). we can guess that water has the structure like DNA or protein or topoisomerases in biology. The DNA-model for water is more probable. This is because that water like DNA has the ability for storing informations.

 We will show that in some conditions, topoisomerase-like waves interact with water, extract the structure of DNA from it and produce life (See figure 37). This is because that the structure of water in extra dimension  includes the structure of DNA. If some waves could reverse the structure of water and replace the structure of water in four dimension by the structure of water in extra dimensions,  they can produce DNA and the life.We can obtain the below equations for waters:

\begin{eqnarray}
&&   H_{water,extra,women}=4\pi T_{tri} \int d\sigma_{11}.. d\sigma_{1}\Sigma_{n.m,X,Y}^{w}\Pi_{a,b,c=1}^{N}\times \nonumber\\ &&[  \Big(P^{water,4-dimension}(a,b,c,n,m,X,Y)H^{water,4-dimension}(a,b,c,n,m,X,Y)\Big)^{-1}\times \nonumber\\ && \Pi_{i=1}^{n}\Pi_{j=1}^{m}\Pi_{k=1}^{X}\Pi_{l=1}^{Y}\delta (\sigma_{nX}^{2}-\theta_{mY}^{2})-\nonumber\\ && P^{DNA,women}(a,b,c,n,m,X,Y) \Pi_{i=1}^{n}Q_{6,un-packed,i,women}\times \nonumber\\&&\Pi_{j=1}^{m}\bar{Q}_{6,packed,j,women}\Pi_{k=1}^{X}E_{6,un-packed,k,women}\Pi_{l=1}^{Y}\bar{E}_{6,packed,l,women}\times\nonumber\\ && \Pi_{i=1}^{n}Q_{5,un-packed,i,women}\Pi_{j=1}^{m}\bar{Q}_{5,packed,j}\times \nonumber\\&&\Pi_{k=1}^{X}E_{5,un-packed,k,women}\Pi_{l=1}^{Y}\bar{E}_{5,packed,l,women}\times\nonumber\\ && \Pi_{i=1}^{n}Q_{3,un-packed,i,women}\Pi_{j=1}^{m}\bar{Q}_{3,packed,j,women}\Pi_{k=1}^{X}E_{3,un-packed,k,women}\Pi_{l=1}^{Y}\bar{E}_{3,packed,l,women}]
\label{wa7}
\end{eqnarray}

\begin{eqnarray}
&&  H_{water,extra, Men}=4\pi T_{tri} \int d\sigma_{11}.. d\sigma_{1}\Sigma_{n.m,X,Y}^{w}\Pi_{a,b,c=1}^{N}\times \nonumber\\ &&[  \Big(P^{water,4-dimension}(a,b,c,n,m,X,Y)H^{water,4-dimension}(a,b,c,n,m,X,Y)\Big)^{-1}\times \nonumber\\ && \Pi_{i=1}^{n}\Pi_{j=1}^{m}\Pi_{k=1}^{X}\Pi_{l=1}^{Y}\delta (\sigma_{nX}^{2}-\theta_{mY}^{2})-\nonumber\\ && P^{DNA, Men}(a,b,c,n,m,X,Y) \Pi_{i=1}^{n}Q_{6,un-packed,i, Men}\Pi_{j=1}^{m}\bar{Q}_{6,packed,j, Men} \times \nonumber\\&&\Pi_{k=1}^{X}E_{6,un-packed,k, Men}\Pi_{l=1}^{Y}\bar{E}_{6,packed,l, Men}\times\nonumber\\ && \Pi_{i=1}^{n}Q_{5,un-packed,i, Men}\Pi_{j=1}^{m}\bar{Q}_{5,packed,j, Men}\Pi_{k=1}^{X}E_{5,un-packed,k, Men}\Pi_{l=1}^{Y}\bar{E}_{5,packed,l, Men}\times\nonumber\\ && \Pi_{i=1}^{n}Q_{3,un-packed,i, Men}\Pi_{j=1}^{m}\bar{Q}_{3,packed,j, Men}\Pi_{k=1}^{X}E_{3,un-packed,k, Men}\Pi_{l=1}^{Y}\bar{E}_{3,packed,l, Men}]
\label{wa8}
\end{eqnarray}

In equations (\ref{wa7},\ref{wa8}), we have shown that the structure of waves and waters have a direct relation with the structure of DNA. Thus, if a wave could reverse the structure of water, the structure of DNA is appeared in four dimension and life is emerged. We have:

\begin{eqnarray}
&&  H_{water,women} + H_{wave,reverse,women}=  H_{DNA, women} 
\label{LI1}
\end{eqnarray}

and

\begin{eqnarray}
&&  H_{water,men} + H_{wave,reverse,men}=  H_{DNA,men}
\label{LI2}
\end{eqnarray}

Solving above equations, we can obtain the Hamiltonian of waves as:

\begin{eqnarray}
&&   H_{wave,reverse,women}=4\pi T_{tri} \int d\sigma_{11}.. d\sigma_{1}\Sigma_{n.m,X,Y}^{w}\Pi_{a,b,c=1}^{N}\times  \nonumber\\ && \Big([ P^{DNA,women}(a,b,c,n,m,X,Y) \Pi_{i=1}^{n}Q_{6,un-packed,i,women}\times \nonumber\\&&\Pi_{j=1}^{m}\bar{Q}_{6,packed,j,women}\Pi_{k=1}^{X}E_{6,un-packed,k,women}\Pi_{l=1}^{Y}\bar{E}_{6,packed,l,women}\times\nonumber\\ && \Pi_{i=1}^{n}Q_{5,un-packed,i,women}\Pi_{j=1}^{m}\bar{Q}_{5,packed,j}\times \nonumber\\&&\Pi_{k=1}^{X}E_{5,un-packed,k,women}\Pi_{l=1}^{Y}\bar{E}_{5,packed,l,women}\times\nonumber\\ && \Pi_{i=1}^{n}Q_{3,un-packed,i,women}\Pi_{j=1}^{m}\bar{Q}_{3,packed,j,women}\Pi_{k=1}^{X}E_{3,un-packed,k,women}\Pi_{l=1}^{Y}\bar{E}_{3,packed,l,women}]-1]|\times \nonumber\\ &&\Pi_{i=1}^{n}\Pi_{j=1}^{m}\Pi_{k=1}^{X}\Pi_{l=1}^{Y}\delta (\sigma_{nX}^{2}-\theta_{mY}^{2})Big)-\nonumber\\ &&[\Big(P^{water,4-dimension}(a,b,c,n,m,X,Y)H^{water,4-dimension}(a,b,c,n,m,X,Y)\Big)^{-1}]\times\nonumber\\ && [P^{DNA,women}(a,b,c,n,m,X,Y) \Pi_{i=1}^{n}Q_{6,un-packed,i,women}\times \nonumber\\&&\Pi_{j=1}^{m}\bar{Q}_{6,packed,j,women}\Pi_{k=1}^{X}E_{6,un-packed,k,women}\Pi_{l=1}^{Y}\bar{E}_{6,packed,l,women}\times\nonumber\\ && \Pi_{i=1}^{n}Q_{5,un-packed,i,women}\Pi_{j=1}^{m}\bar{Q}_{5,packed,j}\times \nonumber\\&&\Pi_{k=1}^{X}E_{5,un-packed,k,women}\Pi_{l=1}^{Y}\bar{E}_{5,packed,l,women}\times\nonumber\\ && \Pi_{i=1}^{n}Q_{3,un-packed,i,women}\Pi_{j=1}^{m}\bar{Q}_{3,packed,j,women}\Pi_{k=1}^{X}E_{3,un-packed,k,women}\Pi_{l=1}^{Y}\bar{E}_{3,packed,l,women}]]
\label{LI3}
\end{eqnarray}

and

\begin{eqnarray}
&&   H_{wave,reverse,men}=4\pi T_{tri} \int d\sigma_{11}.. d\sigma_{1}\Sigma_{n.m,X,Y}^{w}\Pi_{a,b,c=1}^{N}\times  \nonumber\\ && \Big([ P^{DNA,men}(a,b,c,n,m,X,Y) \Pi_{i=1}^{n}Q_{6,un-packed,i,men}\times \nonumber\\&&\Pi_{j=1}^{m}\bar{Q}_{6,packed,j,men}\Pi_{k=1}^{X}E_{6,un-packed,k,men}\Pi_{l=1}^{Y}\bar{E}_{6,packed,l,men}\times\nonumber\\ && \Pi_{i=1}^{n}Q_{5,un-packed,i,men}\Pi_{j=1}^{m}\bar{Q}_{5,packed,j}\times \nonumber\\&&\Pi_{k=1}^{X}E_{5,un-packed,k,men}\Pi_{l=1}^{Y}\bar{E}_{5,packed,l,men}\times\nonumber\\ && \Pi_{i=1}^{n}Q_{3,un-packed,i,men}\Pi_{j=1}^{m}\bar{Q}_{3,packed,j,men}\Pi_{k=1}^{X}E_{3,un-packed,k,men}\Pi_{l=1}^{Y}\bar{E}_{3,packed,l,men}]-1]|\times \nonumber\\ &&\Pi_{i=1}^{n}\Pi_{j=1}^{m}\Pi_{k=1}^{X}\Pi_{l=1}^{Y}\delta (\sigma_{nX}^{2}-\theta_{mY}^{2})Big)-\nonumber\\ &&[\Big(P^{water,4-dimension}(a,b,c,n,m,X,Y)H^{water,4-dimension}(a,b,c,n,m,X,Y)\Big)^{-1}]\times\nonumber\\ && [P^{DNA,men}(a,b,c,n,m,X,Y) \Pi_{i=1}^{n}Q_{6,un-packed,i,men}\times \nonumber\\&&\Pi_{j=1}^{m}\bar{Q}_{6,packed,j,men}\Pi_{k=1}^{X}E_{6,un-packed,k,men}\Pi_{l=1}^{Y}\bar{E}_{6,packed,l,men}\times\nonumber\\ && \Pi_{i=1}^{n}Q_{5,un-packed,i,men}\Pi_{j=1}^{m}\bar{Q}_{5,packed,j}\times \nonumber\\&&\Pi_{k=1}^{X}E_{5,un-packed,k,men}\Pi_{l=1}^{Y}\bar{E}_{5,packed,l,men}\times\nonumber\\ && \Pi_{i=1}^{n}Q_{3,un-packed,i,men}\Pi_{j=1}^{m}\bar{Q}_{3,packed,j,men}\Pi_{k=1}^{X}E_{3,un-packed,k,men}\Pi_{l=1}^{Y}\bar{E}_{3,packed,l,men}]]
\label{LI4}
\end{eqnarray}

Above equations show that for extracting the structure of DNA from pure waters, we need to some special waves. These waves send the structure of waters into extra dimension and recover it's structure from extra dimension. This can be the main reason for the emergence of life.

\begin{figure*}[thbp]
	\begin{center}
		\begin{tabular}{rl}
			\includegraphics[width=8cm]{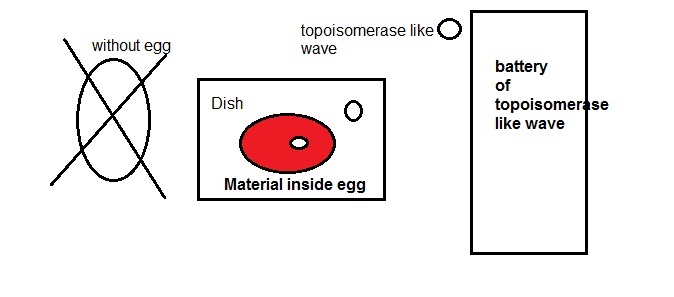}
		\end{tabular}
	\end{center}
	\caption{Producing new born of chicken without needing to egg byy using topoisomerase like waves. }
\end{figure*}

 Now, the question arises that how we could store these main waves . We can show that  there are some mechanism for storing topoisomerase like waves in nature. For example, trees could store all topoisomerase like waves and then apply them for special aims. If we put new born chicken  under some trees, they exchange topoisomerase like waves and trees could act like artificial  mother of these chickens. Using the stored  topoisomerase-like waves, we could control the growth of cells of  fetus of chicken without needing to the egg and produce chicken before 21 days (See figure 38). This can be generalized to human and animals so. 
 
 We have proposed some experimental reasons for programming molecules of two tubes of  water by DNAs of hen and cock (rooster) separately. Then, we have extracted DNAs and  poured   some milk in tubes of water. We have observed that bacteria in milk like lactic acid bacteria interact differently with two types of waters. Then, we have put a metal barrier between two types of water and milk. We expect that electromagnetic waves dont pass this barrier. However, we have observed two different evolutions in milk. This means that some new waves are exchanged between molecules of waters and bacteria in milk. These waves act like topoisomerases in biology and store information in water and DNA.

\begin{figure*}[thbp]
	\begin{center}
		\begin{tabular}{rl}
			\includegraphics[width=8cm]{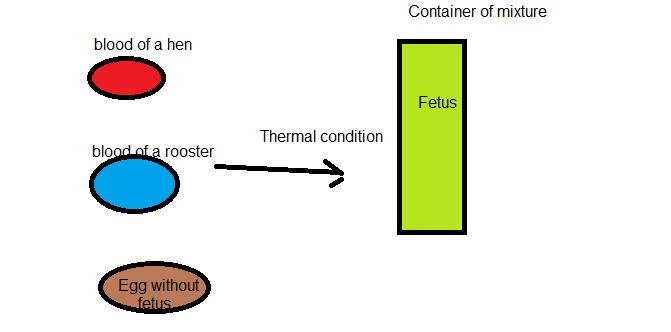}
		\end{tabular}
	\end{center}
	\caption{Producing fetus of chicken by mixing blood cells of a rooster and a hen and returning them to stem ones in matterial of an egg. }
\end{figure*}

Another interesting result is producing a fetus  in material of an egg by mixing stem cells of a hen and cock(Rooster) (See figure 39). We can take blood of a hen and a cock and mix them in some thermal conditions. Then, we  pour material of an egg without any fetus in a container. We  pour mixtures of cells of hen and rooster in that material. After some days, a fetus is emerged in that container.

\section{Summary} \label{sum}
Recently, Montagnier and his collaborators have claimed that a DNA molecule can have communication with other DNA molecules by radiating electromagnetic signals. Using the concepts of string theory and M-theory, we suggested a mathematical model for their idea and calculated the energy and the current  of a normal DNA molecule and a damaged DNA molecule. In our model, each part of a DNA molecule acts like an electrical device. For example, some base pairs act like capacitors and some bases behave like diodes or resistors.  We have discussed that the length of a DNA molecule is much greater than the size of the nucleus and for this reason, a DNA molecule has to be packed in a chromosome. During this packing, at least four types of inductors or coils appear. One coil emerges due to the electric dipoles in the structure of DNA. A second type is produced by coiling DNA bases around the histone in a nucleosome. A third type is created by the formation of loops in a chromatin fiber and finally, a fourth type of inductor emerges along the supercoil within a chromosome.  Each type of these inductors produces one type of magnetic field and plays a main role in a resonant circuit. If we replace bases with electrical devices like inductors and capacitors, we can construct at least four tuned  or  resonant circuits.  Thus, each DNA has at least four  resonant frequencies and many diodes, resistors and antennae. This circuit is very similar to the circuit of an FM radio. We then argued that when a DNA molecule is damaged, its electric charges change and its emitted electromagnetic fields become different respect to the signals of  a normal DNA molecule. For example, if an atom is added to one of the bases, it becomes a heptagonal molecule. If a DNA molecule loses one of its atoms, the hexagonal shape of its base changes and a pentagonal base is produced.  This causes a change in the current  in the DNA's inductors and an extra signal is produced. These signals are received by inductors in other DNA molecules and  an extra current is induced in those DNA molecules, which leads to their destruction. This event may be a reason for the production and replication of cancer cells. If we add a third inductor to this system, which produces an extra current  in the opposite direction with respect to the current of the DNA molecule, an extra magnetic field emerges. This field cancels the effect of the magnetic field which is induced by a damaged DNA molecule in a normal DNA molecule and prevents its destruction.  If we put a damaged DNA molecule of a male or a female near  a damaged DNA of a female or male, their radiated waves cancel the effect of each other and disease progression is stopped.  This is because the types of packing of DNA are different in some chromosomes of men and women and consequently their signals have opposite signs and can cancel the effect of each other in a pair. Finally, the model of this paper lets us to consider the difference in the electromagnetic signals of sperms and eggs. We can show that the radiated waves of sperms which fuse with eggs and produce males are different  respect to the waves of sperms that produce females.

These results can be extended to viruses. One can observe that viruses which are attracted by cells of a female radiate different waves respect to viruses which are absorbed by cells of a male. These viruses are the same, however, type of packing of DNA or the location of chromosomes is different which causes that their signals become different. In fact, some of viruses are male and some others are female. On the other hand, some cancers are produced by viruses. Thus, we can use of the interactions between viruses with different gender to cure cancers. Also, we can construct a wire of viruses that transmit information very fast like wormhole in cosmology or superconductor in nano. These wires includes different type of viruses and can pass the skin and achieve to the special cell.

Another interesting point of viruses is their application as the receiver or sender of electromagnetic waves. We can use of some viruses as the wireless devises for communications with DNAs in various cells. Also, they can help us to communicate with specialized cells and force them to return to initial stem cell or even initial sperm and egg. We can also use of viruses as devices for communications with cells and doing indirect surgery without any touch with human's body.

Above results help us to introduce two types of medical imaging techniques:1.Sperm Medical Imaging (SMI)2.Virus Medical Imaging (VMI). In Sperm Medical Imaging technique, we use of the communications between sperms and cells. In this method, by decoding the radiated signals of sperms and considering  their evolutions, we can imagine the events interior of a human's body.  In Virus Medical Imaging, we use of the communications between Viruses and cells interior of human's body. In this model, by considering the evolutions of viruses out of a body, we can analyse the events interior of a human's body.

To construct a device that communicate with DNAs of cells interior of human's body, we use of the MRI machin which it's protons are replaced by sperms or viruses. In Sperm/Virus Medical Imaging, signals from sperms/Viruses are received by some electronic devices or some special antennas and then applied to the patient. This leads to the occurrence of  a communication  between sperms/Viruses and cells. The DNAs in cells interior of human's body or tissue response to sperms and send  radio frequency waves, which are taken by a radio receiver. These signals may be decoded and the exact information about the evolutions of cells interior of a body can be obtained.

In cancers that have another reason, we can send electromagnetic waves to them and force them to convert to their initial stem cell. Then by the management of their differentiation, we can obtain the non-cancer cell. We can show that in a human's body, two cells in neighbourhood, have reverse signals and cancel the effect of each other. If these waves couldn't remove the effect of each other, the stability of system vanishes and diseases like the cancer are emerged.  

One of interesting subject in biology is the ability of DNA for saving information. We can show that water has the same ability and we can define a coding for it like the genetic coding of DNA. Also, electromagnetic waves have an special type of coding and act similar to DNA or protein. When electromagnetic waves achieve to molecules of water, give their information to water and take it's information. The process is very the same of transcription and translation in biological systems. Sometimes, electromagnetic waves act like DNA and water acts like the protein  and sometimes, reversely, water acts like the DNA and electromagnetic waves act like the protein. On the other hand, water and electromagnetic waves have special genders. For example, molecules of water attract waves with opposite gender.

We can use of the gender of water in medical imaging. First, we separate two waters one with the gender of male which interacts with  men and another with the gender of female which interacts with women. If  a woman eat water wit the gender of female,
 all of it repel with her body. For a pregnant women with a fetus of male, some of this water are absorbed, while for a fetus of female, all of this water go out. This method is completely harmless respect to other methods like the Sonography. Another interesting pint is separating waves by using water. Water with the gender of male attract waves with the gender of female and water with the gender of male, attract waves with the gender of female. Using this property, we can analyse the events interior of human's body by considering the exchanged waves between waters inside calls and outside of body.
 
 In additional to electromagnetic waves, DNAs emit some new types of waves which have the shape of topoisomerases. Topoisomerases are enzymes that participate in the overwinding or underwinding of DNA. We generalise this concept to waves which  participate in the overwinding or underwinding of DNA in extra dimensions (See figure 32).  These wave topoisomerases have 4 + n -dimensional shapes. This is because that DNA  is very compacted in 3 + 1 dimensional space-time. However if we see topology of DNA in extra dimension, we can open packings of it and obtain a simpler shape. In fact DNA acts like a  4 + n-dimensional coil which has been packed to 3 + 1-dimensions. Only, waves could interact with DNA, which could move in 4 +  n dimensions and see the un-packed structure of it in extra dimensions. These waves couldn't be classified as the electromagnetic or gravitational waves. However, by reducing dimensions to 4 or 3, they obtain some properties similar to electromagnetic waves. 
 
 To understand the reason for water memory,we should go to extra dimensions and consider the real structure of water (See figure 33). In fact, water has a coding structure that only part of it can be seen in 4-dimensions. Also,waves have coding structures that only part of their structures can be detected in 4-dimensions. However,by considering amount of memory of water and waves,we can achieve to their real structure in extra dimensions. Also,in extra dimensions,the differences between two genders of water can be cleared more and their structures could be different. Also, an observer in extra dimensions can see two types of gender for waves.Each water can interact with water of opposite gender. Reducing dimensions from 11 or more to 4, some of information about structures of water and waves is lost.
 
One of examples of teleportation via waters is the teleportation of sperms into the  egg (See figure 34). When water of men which includes sperms become near to the water of women which includes egg, they become entangled. When two matters be very entangled, repel other object that be in opposed with this entanglement. For this reason, one of sperms is repelled by waters and transfer into eggs. Another cause of this teleportation is the entanglement between water inside the egg and water outside it. Sperms radiate some topoisomerse-like waves. These waves create some holes in egg. To fill these holes, sperm should be transferred to egg.

Another main subject is the magnetic fields of earth. It seems that magnetic fields of earth are more complicated of what they are seemed and have structures like topoisomerse-like waves. For this reason, they can contribute in teleportation. They produce some holes appropriated with objects that for filling them, objects should be transferred.

Another application of two structures for water in extra dimensions(See figure 35) and waves is radiating two types of  topoisomerase-like waves from earth and sun or other objects in the sky. The origin of these waves are unclear. About earth, it seems that the liquid in it's core produce these types of waves. These waves are needed for activity of cells and growing plants.  Also, these waves help in exchanging information between DNAs and contribute in transcription and translation. About radiated topoisomerase-like waves from sun, we guess that there are some waters in the sky that their waves mix with radiated waves of sun. Although,the structure of sun itself is yet questionable. The radiated waves from sun and earth create a medium that some points of it are holes and empty of fields. To fill these holes, particles could be teleported to them.  In fact,waters of earth and sky are entangled. Each particle enters into these waters, break the entanglement and produces some holes. To fill these holes, particles are teleported into these holes.From this point,we can use for transferring of objects and fields in future (See figure 36).  

 Another question arises that what is the real structure of waters in extra dimensions (See figure 37). we can guess that water has the structure like DNA or protein or topoisomerases in biology. The DNA-model for water is more probable. This is because that water like DNA has the ability for storing informations.
 
 We have shown that in some conditions, topoisomerase-like waves interact with water, extract the structure of DNA from it and produce life. This is because that the structure of water in extra dimension  includes the structure of DNA. If some waves could reverse the structure of water and replace the structure of water in four dimension by the structure of water in extra dimensions,  they can produce DNA and the life.
 
  Now, the question arises that how we could store these main waves. We can show that  there are some mechanism for storing topoisomerase like waves in nature. For example, trees could store all topoisomerase like waves like battery and then apply them for special aims. If we put new born chicken  under some trees, they exchange topoisomerase like waves and trees could act like artificial  mother of these chickens. Using the stored  topoisomerase-like waves, we could control the growth of cells of  fetus of chicken without needing to the egg and produce chicken before 21 days (See figure 38). This can be generalized to human and animals so.  
  
   We have proposed some experimental reasons for programming molecules of two tubes of  water by DNAs of hen and cock (rooster) separately. Then, we have extracted DNAs and  poured   some milk in tubes of water. We have observed that bacteria in milk like lactic acid bacteria interact differently with two types of waters. Then, we have put a metal barrier between two types of water and milk. We expect that electromagnetic waves dont pass this barrier. However, we have observed two different evolutions in milk. This means that some new waves are exchanged between molecules of waters and bacteria in milk. These waves act like topoisomerases in biology and store information in water and DNA.
 
Another interesting result is producing a fetus  in material of an egg by mixing stem cells of a hen and cock(Rooster) (See figure 39). We can take blood of a hen and a cock and mix them in some thermal conditions. Then, we  pour material of an egg without any fetus in a container. We  pour mixtures of specialised cells of hen and rooster in that material. After some days and under some conditions, a fetus is emerged in that container.

\section*{Acknowledgments}
\noindent The work of Alireza Sepehri has been supported financially by Research Institute for biotech development in Tehran.  We are designing a circuit for communication with DNA and we hypothesize that cancer can be cured via this method. We also obtained the first results for producing first stages of fetus of chicken without needing to egg. We are producing a fetus by returning specialized cells of a hen (Chicken) and a rooster to stem ones and then mixing them in material of an egg.  Comments are welcome.

\end{document}